# Decomposed photon anomalous dimension in QCD and the $\{\beta\}$-expanded representations for the Adler function

A. L. Kataev[1,*] and V. S. Molokoedov[1,2,3,4,†]

[1]*Institute for Nuclear Research of the Russian Academy of Science, 117312 Moscow, Russia*
[2]*Research Computing Center, Moscow State University, 119991 Moscow, Russia*
[3]*Moscow Center for Fundamental and Applied Mathematics, 119992 Moscow, Russia*
[4]*Moscow Institute of Physics and Technology, 141700 Dolgoprudny, Moscow Region, Russia*

This work is devoted to the study of the $\{\beta\}$ expansion of the perturbative expressions for the $e^+e^-$ annihilation Adler function $D(Q^2)$ and for the related renormalization group functions, namely, for the photon vacuum polarization function and its anomalous dimension $\gamma(\alpha_s)$ in QCD at the $\mathcal{O}(\alpha_s^4)$ order. We emphasize that $\gamma(\alpha_s)$ is not a conformal-invariant contribution to $D(Q^2)$; therefore, for a consistent analysis, it is necessary to decompose its higher-order perturbation theory coefficients in powers of the $\beta$-function coefficients in the same way as for the Adler function. The arguments in favor of this statement are given. The comparison of the $\overline{\text{MS}}$ and principal of maximum conformality and Brodsky-Lepage-Mackenzie scale approximants are demonstrated. We comment briefly on theoretical and phenomenologically related consequences of this comparison.

## I. INTRODUCTION

Among the various QCD representations for the renormalization-group (RG) invariant quantities studied nowadays are the renormalon-motivated large-$\beta_0$ approximation, discussed in the reviews [1–3], and the $\{\beta\}$-expansion approach, originally proposed in Ref. [4]. It prescribes to decompose the analytically evaluated massless coefficients of the corresponding perturbation theory (PT) QCD series, defined in the gauge-invariant renormalization schemes [such as the minimal-subtraction (MS) like schemes], into the sum of the scale-invariant terms and of the concrete monomials, which contain fixed combinations of the RG $\beta$-function coefficients.

In general, in the fixed order of PT, this representation allows one to go beyond large-$\beta_0$ approximation. Indeed, within the $\{\beta\}$-expansion approach it is possible to trace extra contributions to higher-order PT coefficients from the terms, generated by the subleading renormalon chains. The consequences of the Borel resummations of those ones, proportional to the subleading powers of $\beta_0$, are studied in the recent work of Ref. [5].

The application of the $\{\beta\}$ decomposition is rather useful in the theoretical analysis of the effects of the conformal symmetry and of its violation in the product of the PT expressions for the $e^+e^-$ annihilation Adler function and for the Bjorken polarized sum rule, with their nonsinglet in flavor coefficient functions entering into the Crewther relation [6] and in its QCD generalizations, written down in the form of single power of the $\beta$ function or of multiple power of $\beta$-function representations (see Refs. [7–9] and Ref. [10], respectively). For the studies of the analog of the first of the aforementioned two generalizations of the Crewther relation in the cases of the gauge-dependent renormalization schemes, see Refs. [11,12].

The $\{\beta\}$ decomposition [4] and representations, based on the large-$n_f$ and the large-$\beta_0$ expansions, were applied in the number of formulations of various extensions of the Brodsky-Lepage-Mackenzie (BLM) scale-fixing prescription [13] to higher orders of PT (see [14–17] and [18–20], respectively). Currently, the most popular higher-order BLM extension is the principal of maximum conformality (PMC), which was proposed in Ref. [21].[1] The results of its application essentially depend on a manner of constructing $\{\beta\}$-decomposed coefficients of the specific PT series

---

*kataev@ms2.inr.ac.ru
†viktor_molokoedov@mail.ru

---

[1]The main aim of this scale-fixing approach is to extract the $\{\beta\}$-independent parts of the PT coefficients, which are invariant under the dilatation (but not of the conformal symmetry) transformations. They play a crucial role in the study of the automodality principle of strong interactions [22].

(compare the structure of the expressions presented in Refs. [4,16]).

There are different ways to determine the analytical structure of the $\{\beta\}$-expanded coefficients for the RG-invariant quantities in QCD and QCD-related models.

(1) It is possible to apply the RG-inspired $\mathcal{R}_\delta$ procedure of Ref. [23]. Its definition is rather close to the skeleton-motivated approach proposed in Ref. [24] for determining the energy evolution of the couplings in the analytical PT approach, developed, e.g., in Refs. [25–27].

(2) Another way of fixing the $\{\beta\}$-expansion representation is the multiple $\beta$-function decomposition, suggested in Ref. [28] and developed later on in Refs. [29,30]. It relies considerably on the analogous expansion, introduced in Ref. [10] for the conformal symmetry breaking (CSB) term of the generalized Crewther relation.

It is interesting, but not yet completely understood, that at the three-loop level the application of this approach to the PT series for the static potential [30] reproduces almost all but few coefficients of the $\{\beta\}$-expanded representation fixed through $\mathcal{R}_\delta$ procedure (with taking into account possible misprints in Ref. [31], see discussions in [30]).

(3) There is the approach based on introducing extra degrees of freedom in the gauge model of strong interactions, namely, the Majorana multiplet of massless gluinos. In this scenario, the number of gluinos $n_{\tilde{g}}$ appears in addition to the number of the quark flavors $n_f$ and serves as an auxiliary parameter for fixing terms of the $\{\beta\}$ expansion. This approach was originally proposed in Ref. [4] while considering the three-loop expression for the Adler function in this effective QCD-like model.

It was applied later in Refs. [10,32] at the same order of PT for the case of the Bjorken polarized sum rule.[2] The applicability of this approach to the Adler function and the Bjorken polarized sum rule at the four-loop level was analyzed in Ref. [33].

(4) In the effective QCD-type model with arbitrary numbers of fermion representations at the same PT level, the decomposed expressions for the Adler function and for the Bjorken polarized sum rule were constructed and considered in Refs. [34,35]. The above mentioned studies were based on the results of the analytical calculations of the related RG $\beta$ function, obtained in Ref. [36], and on the expressions given in Ref. [8] for the two important physical quantities mentioned above, analytically evaluated in this QCD-type model at the four-loop level.

---

[2]It should be emphasized that, unlike Ref. [4], where the subsequent BLM procedure was used, in [32] the ideas of the PMC were already realized.

Note that at the next-to-next-to-leading level the expressions of Ref. [34] yield results identical to the ones of Ref. [10], given in another QCD-type model, namely, QCD supplemented by the multiplet of massless gluinos.

Apart from the ambiguities in constructing the $\{\beta\}$-expanded coefficients of the RG-invariant functions, there is still no consensus on how to apply the PMC approach to the physically important quantities, namely, to the $e^+e^-$ annihilation Adler function [and to the associated $R_{e^+e^-}(s)$ and $R_\tau$ ratios] and to the Bjorken polarized sum rule (which has the PT expression closely related to the one of the Gross–Llewellyn-Smith sum rule of the deep-inelastic $\nu N$ scattering [37,38]).

Indeed, despite the existence of the arguments presented in Refs. [29,32,39] that the original PMC analysis, performed in Refs. [23,31], for instance, for the Adler function, has definite drawbacks related to the absence of the $\{\beta\}$ expansion for the anomalous dimension of the photon vacuum polarization function, in further PMC-based works, including the most recent ones (see, e.g., [40–43]), these arguments were not accepted.

Note that the renormalization group method, elaborated in the classical works of Refs. [44–46], was applied in Ref. [47], where its relation to the principle of minimal sensitivity (PMS), formulated in the work of Ref. [48], was clarified. The theoretical basis of this approach was questioned several times (see, e.g., Refs. [43,49,50]). We will leave aside here the possible scientific study of the published critics of PMS presented in these works and definite arguments in favor of the PMS defense (see, e.g., [51]).

In our work, we will concentrate on the more detailed analysis of the problem associated with consistent, from our point of view, application of the PMC approach primarily to the Adler function $D(Q^2)$. After presenting technical details in Sec. II, in Sec. III we will demonstrate theoretical consequences of taking into account the $\{\beta\}$-expanded representation for the QCD expression for the anomalous dimension $\gamma(\alpha_s)$ of the photon vacuum polarization function $\Pi(Q^2)$, which we will call the photon anomalous dimension. In Sec. IV, we will provide extra theoretical arguments in favor of the necessity of applying $\{\beta\}$ decomposition for $\gamma(\alpha_s)$.

We shall clarify that only after this step is it possible to understand where the important parts of the renormalon contributions to the Adler function, studied in Refs. [1,2,52], were put in shadow in the previous PMC-related considerations.

In Sec. V, we construct a properly defined by us new Adler PMC-type approximation related to taking into account the truncated next-to-leading order (NLO), next-to-next-to-leading order (N$^2$LO) and next-to-next-to-next-to-leading order (N$^3$LO) $\overline{\text{MS}}$ massless PT QCD expressions. In part, the qualitative phenomenological comparison with the analogous $\overline{\text{MS}}$-scheme approximations will be presented.

The explicit solutions of the RG equations for functions $\Pi(Q^2)$ and $D(Q^2)$ are given in Appendix A. In Appendix B, the coefficients of the expansion of the Adler function in powers of the $\beta$-function coefficients with and without taking into account the application of the $\{\beta\}$ decomposition to $\gamma(\alpha_s)$ are presented.

## II. PRELIMINARIES

Let us start from the consideration of the $e^+e^-$ annihilation Adler function. It is defined in the Euclidean region as

$$D(L, a_s) = -\frac{d\Pi(L, a_s)}{d \ln Q^2} = Q^2 \int_0^\infty ds \frac{R_{e^+e^-}(l, a_s)}{(s + Q^2)^2}, \quad (1)$$

where $a_s(\mu^2) = a_s = \alpha_s/\pi$, $\alpha_s$ is the $\overline{\text{MS}}$-scheme strong coupling constant, $\mu$ is the renormalization scale, $L = \ln(\mu^2/Q^2)$ and $l = \ln(\mu^2/s)$, respectively, $Q^2 = -q^2 > 0$ is the Euclidean kinematic variable, and $s = q^2 > 0$ is the timelike Minkowskian variable.

The spectral function $R_{e^+e^-}(l, a_s)$ is the theoretical expression for the electron-positron annihilation $R$ ratio. It is proportional to the experimentally measured total cross section $\sigma(e^+e^- \to \gamma^* \to \text{hadrons})$. $\Pi(L, a_s)$ is the renormalized QCD expression for the photon vacuum polarization function, which enters in the two-point correlator $\Pi_{\mu\nu}(q)$ of the electromagnetic quark vector currents $j^\mu(x)$ as

$$\Pi_{\mu\nu}(q) = i \int d^4 x \, e^{iqx} \langle 0 | T\{j_\mu(x) j_\nu(0)\} | 0 \rangle$$
$$= \frac{1}{12\pi^2}(q_\mu q_\nu - q^2 g_{\mu\nu})\Pi(q^2). \quad (2)$$

Here $j^\mu(x) = \sum_f Q_f \bar{\psi}_f(x) \gamma^\mu \psi_f(x)$, and $Q_f$ stands for the electric charge of the quark field $\psi_f(x)$ with flavor $f$. Note that, since the vector current is conserved both in the renormalized and bare cases, the expression for the tensor $\Pi_{\mu\nu}(q)$ is transverse in both cases as well.

The detailed theoretical studies, conducted in Refs. [53,54] and used later on in Refs. [55,56], lead to the following renormalization prescription for the photon vacuum polarization function in QCD:

$$\Pi(L, a_s) = Z(a_s) + \Pi_B(L, a_{sB}). \quad (3)$$

Here $\Pi_B(L, a_{sB})$ is the unrenormalized QCD expression for the vacuum polarization function; $a_{sB} = \alpha_{sB}/\pi = \mu^{2\varepsilon} Z_{a_s}(a_s) a_s$, $\alpha_{sB}$ is the bare strong coupling, and $Z_{a_s}(a_s)$ is the corresponding renormalization constant, which defines the QCD RG $\beta$ function. $Z(a_s) = (Z_{ph}(a_s) - 1)/a$, where $a = \alpha/\pi$ and $\alpha$ is the renormalized QED coupling, defined in the case where the effects of its QED running are not taken into account. $Z_{ph}(a_s)$ is the renormalization constant of the photon wave function, considered in the case where the QCD corrections only are taken into account.

Within the class of the MS-like subtraction schemes, the expression for $Z(a_s)$ contains the pole terms in $\varepsilon$,

$$Z(a_s) = \sum_{p \geq 1} a_s^{p-1} \sum_{k=1}^p \frac{Z_{p,-k}}{\varepsilon^k}, \quad (4)$$

whereas the quantity $\Pi_B(L, a_{sB})$ has the following form:

$$\Pi_B(L, a_{sB}) = \sum_{p \geq 1} \left(\frac{\mu^2}{Q^2}\right)^{\varepsilon p} a_{sB}^{p-1} \sum_{k=-p}^\infty \Pi_{p,k} \varepsilon^k, \quad (5)$$

where $\varepsilon = (4 - d)/2$ and $d$ is the space-time dimension.

The renormalized photon vacuum polarization function $\Pi(L, a_s)$ obeys the following inhomogeneous RG equation:

$$\left(\frac{\partial}{\partial \ln \mu^2} + \beta(a_s) \frac{\partial}{\partial a_s}\right) \Pi(L, a_s) = \gamma(a_s), \quad (6)$$

where

$$\gamma(a_s) = \frac{d\Pi(L, a_s)}{d \ln \mu^2} \quad (7)$$

is the QCD anomalous dimension of $\Pi(L, a_s)$ and $\beta(a_s)$ defines the scale dependence of the strong coupling RG $\beta$ function, namely,

$$\beta(a_s) = \frac{\partial a_s}{\partial \ln \mu^2} = -\sum_{n \geq 0} \beta_n a_s^{n+2}. \quad (8)$$

One should note that the RG $\beta$ function is included in the renormalized expression for the trace of the energy-momentum tensor in the massless QCD; therefore, it is a measure of violation not only of the symmetry under the dilatation transformations, but under the conformal ones as well.

Application of the renormalization procedure leads to the following perturbative expression for the photon vacuum polarization function:

$$\Pi(L, a_s) = d_R \left(\sum_f Q_f^2\right) \Pi^{NS}(L, a_s) + d_R \left(\sum_f Q_f\right)^2 \Pi^{SI}(L, a_s), \quad (9)$$

where $d_R$ is the dimension of the fundamental representation of the considered generic simple gauge group. In our study, we are primarily interested in the case of the $SU(N_c)$ gauge group with $d_R = N_c$. In its particular case of the $SU(3)$ color group, relevant for physical QCD, $N_c = 3$. The quantities $\Pi^{NS}(L, a_s)$ and $\Pi^{SI}(L, a_s)$ are the flavor nonsinglet (NS) and singlet (SI) contributions to $\Pi(L, a_s)$, respectively.

Substituting $\Pi(L, a_s)$ from Eq. (9) into (6), one can get the PT expression for the photon anomalous dimension,

$$\gamma(a_s) = d_R\left(\sum_f Q_f^2\right)\gamma^{NS}(a_s) + d_R\left(\sum_f Q_f\right)^2 \gamma^{SI}(a_s), \quad (10)$$

where $\gamma^{NS}(a_s)$ and $\gamma^{SI}(a_s)$ are the PT series in strong coupling,

$$\gamma^{NS}(a_s) = \sum_{n\geq 0}\gamma_n a_s^n, \qquad \gamma^{SI}(a_s) = \sum_{n\geq 3}\gamma_n^{SI} a_s^n. \quad (11)$$

Taking into account Eqs. (6) and (8), one arrives at the following RG-improved expressions for $\Pi^{NS}(L, a_s)$ and $\Pi^{SI}(L, a_s)$ at $L = 0$:

$$\Pi^{NS}(0, a_s(Q^2)) = \sum_{n\geq 0}\Pi_n a_s^n(Q^2),$$

$$\Pi^{SI}(0, a_s(Q^2)) = \sum_{n\geq 3}\Pi_n^{SI} a_s^n(Q^2). \quad (12)$$

The solution of the RG equation (6) can be found perturbatively. Its explicit form obtained at the $\mathcal{O}(a_s^4)$ level is presented in Appendix A.

Using the expressions presented above, it is possible to derive the following expression for the Adler function:

$$D(L, a_s) = \gamma(a_s) - \beta(a_s)\frac{\partial}{\partial a_s}\Pi(L, a_s). \quad (13)$$

In contrast to the polarization operator, it is the RG-invariant quantity. Therefore, it satisfies the homogeneous RG equation,

$$\frac{dD(L, a_s)}{d\ln\mu^2} = \left(\frac{\partial}{\partial \ln\mu^2} + \beta(a_s)\frac{\partial}{\partial a_s}\right)D(L, a_s) = 0. \quad (14)$$

Solving the system of the corresponding RG equations, one can get the following PT expression for the Adler function:

$$D(a_s(Q^2)) = d_R\left(\sum_f Q_f^2\right)D^{NS}(a_s(Q^2)) + d_R\left(\sum_f Q_f\right)^2 D^{SI}(a_s(Q^2)), \quad (15)$$

where its NS and SI contributions are defined as

$$D^{NS}(a_s(Q^2)) = \sum_{n\geq 0}d_n a_s^n(Q^2), \quad (16a)$$

$$D^{SI}(a_s(Q^2)) = \sum_{n\geq 3}d_n^{SI} a_s^n(Q^2). \quad (16b)$$

In the massless limit, all logarithmic corrections to $D(Q^2)$, controlled by the RG, can be summed up into the running coupling $a_s(Q^2)$.

Using the explicit solution of Eq. (6) for $\Pi(L, a_s)$, one can obtain the solution of the RG equation (14), expressed in terms of the PT coefficients of $\Pi(L, a_s)$, $\beta(a_s)$, and $\gamma(a_s)$. The explicit form of its $\mathcal{O}(a_s^4)$ approximation is given in Appendix A as well.

Comparing solutions of the expressions of Eqs. (16a) and (16b) with the ones following from Eq. (13) and taking into account the dependence $a_s(Q^2)$ on the Euclidean momentum $Q^2$, we can obtain the following relations:

$$d_0 = \gamma_0, \quad (17a)$$

$$d_1 = \gamma_1, \quad (17b)$$

$$d_2 = \gamma_2 + \beta_0\Pi_1, \quad (17c)$$

$$d_3 = \gamma_3 + 2\beta_0\Pi_2 + \beta_1\Pi_1, \quad (17d)$$

$$d_4 = \gamma_4 + 3\beta_0\Pi_3 + 2\beta_1\Pi_2 + \beta_2\Pi_1, \quad (17e)$$

$$d_3^{SI} = \gamma_3^{SI}, \quad (17f)$$

$$d_4^{SI} = \gamma_4^{SI} + 3\beta_0\Pi_3^{SI}. \quad (17g)$$

One should recall that, in the class of the gauge-invariant MS-like schemes, the scheme dependence of the coefficients $d_k$ starts to manifest itself from $k \geq 2$ due to scheme-dependent terms $\Pi_m$ at $m \geq 1$. The expressions (17a)–(17g) are derived comparing the RG-based relation directly associating the Adler function to the photon vacuum polarization function and to its anomalous dimension.

The analytical expressions for the coefficients $d_0 \div d_4$ and $\gamma_0 \div \gamma_4$, $\Pi_0 \div \Pi_3$ may be found in Refs. [57,58] correspondingly (see also references therein).

In the MS-like scheme, the coefficients of the corresponding RG $\beta$ function up to the $\beta_3$ term are known from the results of the analytical calculations of Ref. [59], which were effectively confirmed by the foundation of nullification of the three-loop $DR$-like scheme approximation for the RG $\beta$ function of $\mathcal{N} = 4$ supersymmetry (SUSY) Yang-Mills theory (see, e.g., [60]) and by the direct analytical QCD calculations from Ref. [61].

## III. THE $\{\beta\}$ EXPANSION OF $\gamma(a_s)$ AND $\Pi(a_s)$

As was already mentioned in the Introduction, the $\{\beta\}$-expansion formalism implies the representation of the expressions of higher-order PT corrections to the massless RG-invariant quantities, evaluated in the gauge-invariant renormalization schemes, through the sums of the monomials in powers of the RG $\beta$-function coefficients by separating the scale-invariant contributions.

For instance, the coefficients $d_1 \div d_4$ and $d_3^{\mathrm{SI}} \div d_4^{\mathrm{SI}}$ of the $e^+e^-$ annihilation Adler function, defined in Eqs. (15)–(16b), have the following $\{\beta\}$-expanded structure:

$$d_1 = d_1[0], \tag{18a}$$

$$d_2 = \beta_0 d_2[1] + d_2[0], \tag{18b}$$

$$d_3 = \beta_0^2 d_3[2] + \beta_1 d_3[0,1] + \beta_0 d_3[1] + d_3[0], \tag{18c}$$

$$d_4 = \beta_0^3 d_4[3] + \beta_1 \beta_0 d_4[1,1] + \beta_2 d_4[0,0,1] + \beta_0^2 d_4[2] + \beta_1 d_4[0,1] + \beta_0 d_4[1] + d_4[0], \tag{18d}$$

$$d_3^{\mathrm{SI}} = d_3^{\mathrm{SI}}[0], \tag{18e}$$

$$d_4^{\mathrm{SI}} = \beta_0 d_4^{\mathrm{SI}}[1] + d_4^{\mathrm{SI}}[0], \tag{18f}$$

where terms $d_k[\ldots]$, $d_k^{\mathrm{SI}}[\ldots]$ do not contain $n_f$ dependence (except for the flavor $n_f$ dependence arising from the contributions to $d_4[0]$ of the light-by-light scattering-type diagrams [28,29,34]).

Starting from $k \geq 3$, there is an ambiguity associated with fixing the terms $d_k[\ldots]$ (apart from the coefficients $d_k[k-1]$ defined by the leading renormalon chains effects, obtained in the case of QED in Ref. [62] and reformulated to the case of QCD in Ref. [7]). It is related to the differences of splitting of the $n_f$-dependent contributions to higher-order corrections to RG-invariant quantity between flavor-dependent coefficients $\beta_i$ (with $i \geq 0$) of the RG $\beta$ function (see, e.g., [4,32,39]).

One of the currently existing ways to resolve this problem was proposed in Ref. [28] and more widely studied in Refs. [29,30]. In accordance with these considerations, the PT expression for the NS contribution to the Adler function at the $\mathcal{O}(a_s^4)$ level can be presented through the following double sum:

$$\begin{aligned}D^{\mathrm{NS}}(a_s(Q^2)) &= 1 + \sum_{n=0}^{3}\left(\frac{\beta(a_s(Q^2))}{a_s(Q^2)}\right)^n \sum_{k=1}^{4-n} D_{n,k} a_s^k(Q^2) \\ &= 1 + D_{0,1} a_s(Q^2) + (D_{0,2} - \beta_0 D_{1,1}) a_s^2(Q^2) + (D_{0,3} - \beta_0 D_{1,2} - \beta_1 D_{1,1} + \beta_0^2 D_{2,1}) a_s^3(Q^2) \\ &\quad + (D_{0,4} - \beta_0 D_{1,3} - \beta_1 D_{1,2} - \beta_2 D_{1,1} + \beta_0^2 D_{2,2} + 2\beta_0 \beta_1 D_{2,1} - \beta_0^3 D_{3,1}) a_s^4(Q^2).\end{aligned} \tag{19}$$

Here, we will not touch upon the grounds of those presented in Eq. (19)-type representations. The arguments in its favor are given in Refs. [28–30]. In fact, the theoretical ways of fixing analytical expressions of the terms $d_k[\ldots]$ are not unique (see, e.g., definite considerations presented in Refs. [29,34,35]). Here we will consider the one outlined in Ref. [28] and followed in Ref. [30].

As observed by us there, it is applicable to the wider class of the functions and quantities in the corresponding RG equations. One may ask the following question: what are the theoretical and phenomenological reasons for separating the scale-invariant contributions $d_k[0]$ from the total expressions for the coefficients $d_k$?

The problem of careful consideration of the status of the PMC-related expressions and of the unraveling in them of the effects related to the scale-invariant limit and to its violation by the CSB effects are among the answers to this question.

The PMC-related considerations enable one to eliminate $\beta$-dependent terms in the coefficients $d_k$ of Eqs. (18b)–(18d) and (18f) by redefining the scale parameter in every order of PT and to leave in the coefficients of the PT expressions related to the Green's function quantities the scale-invariant parts $d_k[0]$ only.

As a result, the scale parameter becomes the coupling-dependent function (for the concrete realization of this feature within large-$n_f$ expansion, see Refs. [14–16]; its PMC-type realizations are given in Refs. [21,32]). It is also important that the PT coefficients of higher-order corrections to the corresponding physical quantities, studied in the gauge-invariant schemes, are becoming independent on the choice of scale.

The representation of Eq. (19) allows one not only to separate the scale-invariant contributions $d_k[0]$, but to reproduce the structure of the $\{\beta\}$ expansion in Eqs. (18b)–(18d) as well. It also imposes essential restrictions on the terms of this decomposition, namely,

$$d_2[1] = d_3[0,1] = d_4[0,0,1] = -D_{1,1}, \tag{20a}$$

$$d_3[1] = d_4[0,1] = -D_{1,2}, \tag{20b}$$

$$d_3[2] = d_4[1,1]/2 = D_{2,1}. \tag{20c}$$

This property is in correspondence with the feature of "special degeneracy" observed in Ref. [23] while applying the considered $\mathcal{R}_\delta$ procedure.

Application of these relations allowed the authors of Ref. [28] to get the analytical expressions for the terms $d_k[\ldots]$ with $k \leq 4$ in the $\{\beta\}$ expansions (18b)–(18d). Their explicit form is given in Appendix B.

The representation (19) also enabled us to fix several terms of the $\{\beta\}$ expansion of the totally unknown, at present, coefficient $d_5$ [29].

In the approach described above, for finding terms $d_k[\ldots]$ of the $\{\beta\}$-decomposed corrections to the $e^+e^-$ annihilation Adler function, it is not necessary to use any information about the possible $\{\beta\}$ structure of the RG-related quantities, such as the photon anomalous dimension or the vacuum polarization function. In this case, we deal directly with the RG-invariant quantity $D(Q^2)$. However, when we pass to consideration of the relation (13) between $D(L, a_s)$, $\gamma(a_s)$, and $\Pi(L, a_s)$ and to the expressions (17a)–(17g) following from it, the important issue of whether or not to decompose the coefficients of $\gamma(a_s)$ and $\Pi(L, a_s)$ vividly arises. We adhere here to the statement made previously in Refs. [29,32,39] that it is really necessary to decompose them in combinations of the $\beta$-function coefficients. In accordance with this opinion, in order to extract the scale-invariant contributions from the PT expressions for the photon anomalous dimension, we should apply the $\{\beta\}$-expansion procedure to the coefficients $\gamma_k$ and $\gamma_k^{SI}$ of the photon anomalous dimension function $\gamma(a_s)$ in Eq. (11) as well. Additional arguments in favor of this assertion will be given below.

Following the proposal of Ref. [39], we write

$$\gamma_1 = \gamma_1[0], \tag{21a}$$

$$\gamma_2 = \beta_0 \gamma_2[1] + \gamma_2[0], \tag{21b}$$

$$\gamma_3 = \beta_0^2 \gamma_3[2] + \beta_1 \gamma_3[0,1] + \beta_0 \gamma_3[1] + \gamma_3[0], \tag{21c}$$

$$\gamma_4 = \beta_0^3 \gamma_4[3] + \beta_1 \beta_0 \gamma_4[1,1] + \beta_2 \gamma_4[0,0,1] + \beta_0^2 \gamma_4[2] + \beta_1 \gamma_4[0,1] + \beta_0 \gamma_4[1] + \gamma_4[0], \tag{21d}$$

$$\gamma_3^{SI} = \gamma_3^{SI}[0], \tag{21e}$$

$$\gamma_4^{SI} = \beta_0 \gamma_4^{SI}[1] + \gamma_4^{SI}[0]. \tag{21f}$$

Equation (13) leads to the relations $d_k[0] = \gamma_k[0]$ and $d_k^{SI}[0] = \gamma_k^{SI}[0]$. This fact, in conjunction with the $\{\beta\}$ expansion (18a)–(18d) and equalities (20a)–(20c), entails the following relationships for terms $\gamma_k[\ldots]$ of the photon anomalous dimension:

$$\gamma_2[1] = \gamma_3[0,1] = \gamma_4[0,0,1], \quad \gamma_3[1] = \gamma_4[0,1],$$
$$\gamma_3[2] = \gamma_4[1,1]/2. \tag{22}$$

Accommodating the explicit analytical expressions for the coefficients $\gamma_0 \div \gamma_4$ [58] and $\beta_0 \div \beta_2$ [59] and using the relations (22), we obtain the analytical expressions for terms $\gamma_k[\ldots]$ and $\gamma_k^{SI}[\ldots]$,

$$\gamma_1[0] = \frac{3}{4} C_F, \quad \gamma_2[0] = -\frac{3}{32} C_F^2 + \frac{1}{16} C_F C_A, \quad \gamma_2[1] = \gamma_3[0,1] = \gamma_4[0,0,1] = \frac{11}{16} C_F, \tag{23a}$$

$$\gamma_3[1] = \gamma_4[0,1] = \left(\frac{239}{192} - \frac{11}{4} \zeta_3\right) C_F^2 + \left(\frac{163}{288} + \frac{11}{4} \zeta_3\right) C_F C_A, \quad \gamma_3[2] = \frac{1}{2} \gamma_4[1,1] = -\frac{77}{144} C_F, \tag{23b}$$

$$\gamma_3[0] = -\frac{69}{128} C_F^3 + \left(-\frac{101}{256} + \frac{33}{16} \zeta_3\right) C_F^2 C_A + \left(-\frac{53}{192} - \frac{33}{16} \zeta_3\right) C_F C_A^2, \tag{23c}$$

$$\gamma_4[2] = \left(\frac{5467}{1536} - \frac{119}{16} \zeta_3 + \frac{99}{32} \zeta_4\right) C_F^2 + \left(-\frac{123}{512} + \frac{629}{64} \zeta_3 - \frac{99}{32} \zeta_4\right) C_F C_A, \tag{23d}$$

$$\gamma_4[1] = \left(-\frac{1477}{256} - \frac{135}{32} \zeta_3 + \frac{435}{32} \zeta_5\right) C_F^3 + \left(\frac{4733}{2048} + \frac{1167}{128} \zeta_3 - \frac{297}{128} \zeta_4 - \frac{765}{64} \zeta_5\right) C_F^2 C_A$$
$$+ \left(-\frac{16453}{18432} - \frac{2109}{256} \zeta_3 + \frac{297}{128} \zeta_4 - \frac{135}{128} \zeta_5\right) C_F C_A^2, \quad \gamma_4[3] = \left(-\frac{107}{384} - \frac{3}{8} \zeta_3\right) C_F, \tag{23e}$$

$$\gamma_4[0] = \left(\frac{4157}{2048} + \frac{3}{8} \zeta_3\right) C_F^4 + \left(-\frac{3509}{1536} - \frac{73}{128} \zeta_3 - \frac{165}{32} \zeta_5\right) C_F^3 C_A + \left(\frac{9181}{4608} + \frac{299}{128} \zeta_3 + \frac{165}{64} \zeta_5\right) C_F^2 C_A^2$$
$$+ \left(-\frac{30863}{36864} - \frac{147}{128} \zeta_3 + \frac{165}{64} \zeta_5\right) C_F C_A^3 + \left(\frac{3}{16} - \frac{1}{4} \zeta_3 - \frac{5}{4} \zeta_5\right) \frac{d_F^{abcd} d_A^{abcd}}{d_R} + \left(-\frac{13}{16} - \zeta_3 + \frac{5}{2} \zeta_5\right) \frac{d_F^{abcd} d_F^{abcd}}{d_R} n_f, \tag{23f}$$

$$\gamma_3^{\text{SI}}[0] = \left(\frac{11}{192} - \frac{1}{8}\zeta_3\right)\frac{d^{abc}d^{abc}}{d_R}, \qquad \gamma_4^{\text{SI}}[1] = \left(\frac{55}{256} - \frac{123}{256}\zeta_3 + \frac{9}{64}\zeta_4 + \frac{15}{64}\zeta_5\right)\frac{d^{abc}d^{abc}}{d_R}, \quad (23\text{g})$$

$$\gamma_4^{\text{SI}}[0] = \left(\left(-\frac{13}{64} - \frac{1}{4}\zeta_3 + \frac{5}{8}\zeta_5\right)C_F + \left(\frac{205}{1536} - \frac{13}{64}\zeta_3 - \frac{5}{32}\zeta_5\right)C_A\right)\frac{d^{abc}d^{abc}}{d_R}, \quad (23\text{h})$$

where $\zeta_n = \sum_{k\geq 1}^{\infty} k^{-n}$ is the Riemann $\zeta$ function; $C_F$ and $C_A$ are the quadratic Casimir operators in the fundamental and adjoint representation of the gauge group correspondingly, $d^{abc} = 2\text{Tr}(t^a t^{\{b} t^{c\}})$, $d_F^{abcd} = \text{Tr}(t^a t^{\{b} t^c t^{d\}})/6$, and $d_A^{abcd} = \text{Tr}(C^a C^{\{b} C^c C^{d\}})/6$, $(C^a)_{bc} = -if^{abc}$ are the generators of the adjoint representation with the antisymmetric structure constants $f^{abc}$: $[t^a, t^b] = if^{abc}t^c$. The terms proportional to $d_F^{abcd}d_F^{abcd}n_f/d_R$ and $d_F^{abcd}d_A^{abcd}/d_R$ group structures are the light-by-light scattering effects and they have to be included in the scale-invariant "$n_f$-independent" coefficient $\gamma_4[0]$ [28,29].

Using now the expressions (17a)–(17g) and (21a)–(21f) and taking into account the following $\{\beta\}$-expansion structure of the vacuum polarization function (9) and (12),

$$\Pi_0 = \Pi_0[0], \qquad \Pi_1 = \Pi_1[0], \qquad \Pi_2 = \Pi_2[0] + \beta_0 \Pi_2[1], \qquad \Pi_3 = \Pi_3[0] + \beta_0 \Pi_3[1] + \beta_1 \Pi_3[0,1] + \beta_0^2 \Pi_3[2], \quad (24)$$

we arrive at the substantial relationships between terms $d_k[\ldots]$, $\gamma_k[\ldots]$, and $\Pi_k[\ldots]$,

$$d_1[0] = \gamma_1[0], \qquad d_2[0] = \gamma_2[0], \qquad d_2[1] = \gamma_2[1] + \Pi_1[0], \quad (25\text{a})$$

$$d_3[0] = \gamma_3[0], \qquad d_3[1] = \gamma_3[1] + 2\Pi_2[0], \qquad d_3[0,1] = \gamma_3[0,1] + \Pi_1[0], \qquad d_3[2] = \gamma_3[2] + 2\Pi_2[1], \quad (25\text{b})$$

$$d_4[0] = \gamma_4[0], \qquad d_4[1] = \gamma_4[1] + 3\Pi_3[0], \qquad d_4[0,1] = \gamma_4[0,1] + 2\Pi_2[0], \qquad d_4[2] = \gamma_4[2] + 3\Pi_3[1], \quad (25\text{c})$$

$$d_4[3] = \gamma_4[3] + 3\Pi_3[2], \qquad d_4[1,1] = \gamma_4[1,1] + 3\Pi_3[0,1] + 2\Pi_2[1], \quad (25\text{d})$$

$$d_4[0,0,1] = \gamma_4[0,0,1] + \Pi_1[0], \qquad d_3^{\text{SI}}[0] = \gamma_3^{\text{SI}}[0], \qquad d_4^{\text{SI}}[0] = \gamma_4^{\text{SI}}[0], \qquad d_4^{\text{SI}}[1] = \gamma_4^{\text{SI}}[1] + 3\Pi_3^{\text{SI}}[0], \quad (25\text{e})$$

where $\Pi_3^{\text{SI}}[0] = \Pi_3^{\text{SI}}$. Applying these relations and using the explicit expressions for $d_k[\ldots]$ from Refs. [28,29] and for $\gamma_k[\ldots]$ from Eqs. (23a)–(23h), we get the values of terms $\Pi_k[\ldots]$,

$$\Pi_0[0] = \frac{5}{3}, \qquad \Pi_1[0] = \left(\frac{55}{16} - 3\zeta_3\right)C_F, \qquad \Pi_2[1] = \left(\frac{3701}{288} - \frac{19}{2}\zeta_3\right)C_F, \quad (26\text{a})$$

$$\Pi_2[0] = \left(-\frac{143}{96} - \frac{37}{8}\zeta_3 + \frac{15}{2}\zeta_5\right)C_F^2 + \left(\frac{73}{72} - \frac{3}{4}\zeta_3 - \frac{5}{4}\zeta_5\right)C_F C_A, \quad (26\text{b})$$

$$\Pi_3[2] = \left(\frac{196513}{3456} - \frac{809}{24}\zeta_3 - 15\zeta_5\right)C_F, \qquad \Pi_3[0,1] = \left(\frac{3701}{432} - \frac{19}{3}\zeta_3\right)C_F, \quad (26\text{c})$$

$$\Pi_3[1] = \left(-\frac{22103}{4608} - \frac{1439}{24}\zeta_3 + 9\zeta_3^2 - \frac{33}{32}\zeta_4 + \frac{125}{2}\zeta_5\right)C_F^2 + \left(\frac{29353}{1536} - \frac{473}{192}\zeta_3 - \frac{3}{2}\zeta_3^2 + \frac{33}{32}\zeta_4 - \frac{185}{12}\zeta_5\right)C_F C_A, \quad (26\text{d})$$

$$\Pi_3[0] = \left(-\frac{31}{256} + \frac{39}{32}\zeta_3 + \frac{735}{32}\zeta_5 - \frac{105}{4}\zeta_7\right)C_F^3 + \left(-\frac{520933}{55296} + \frac{5699}{384}\zeta_3 - \frac{33}{4}\zeta_3^2 + \frac{99}{128}\zeta_4 - \frac{565}{64}\zeta_5 + \frac{105}{8}\zeta_7\right)C_F^2 C_A$$
$$- \left(\frac{112907}{55296} + \frac{5839}{768}\zeta_3 - \frac{33}{4}\zeta_3^2 + \frac{99}{128}\zeta_4 - \frac{835}{384}\zeta_5 + \frac{35}{16}\zeta_7\right)C_F C_A^2, \quad (26\text{e})$$

$$\Pi_3^{\text{SI}} = \Pi_3^{\text{SI}}[0] = \left(\frac{431}{2304} - \frac{63}{256}\zeta_3 - \frac{1}{8}\zeta_3^2 - \frac{3}{64}\zeta_4 + \frac{15}{64}\zeta_5\right)\frac{d^{abc}d^{abc}}{d_R}. \quad (26\text{f})$$

One should note that, in contrast to Eqs. (20a) and (22), for the analogous $\{\beta\}$-expanded terms of the photon vacuum polarization function, one has $\Pi_3[0,1] \neq \Pi_2[1]$. However, they turn out to be proportional to each other, namely, $\Pi_3[0,1] = 2/3 \cdot \Pi_2[1]$. This follows from the fact that the derivative $\partial/\partial a_s$ is included in Eq. (13). Thus, the analog of the double sum representation (19) for the two-point correlator $\Pi(a_s)$ is not fulfilled, but is held for the term $\beta(a_s)\partial\Pi(a_s)/\partial a_s$ in Eq. (13) as the whole.

After receiving the concrete results (23a)–(23h) and (26a)–(26f) for the $\{\beta\}$-expanded coefficients of $\gamma(a_s)$ and $\Pi(a_s)$, respectively, we now move on to presenting extra arguments in favor of the necessity of their $\{\beta\}$ decomposition.

## IV. ARGUMENTS IN FAVOR OF THE $\{\beta\}$ EXPANSION OF $\gamma(a_s)$

While realizing the PMC ideas to the $\overline{\mathrm{MS}}$-scheme expression for the quantities related to the Adler function, expressed in Refs. [16,21], the authors of Refs. [23,31] adhere to the point of view that the $\{\beta\}$-expansion procedure should not be applied to the photon anomalous dimension $\gamma(a_s)$, which enters in the presented above Eq. (13).

In the work of Ref. [32] within the effective QCD model with multiplet of massless gluinos, the attempt to clarify that the nonapplication of the $\{\beta\}$-expansion approach to $\gamma(a_s)$ contradicts the renormalizability principles was made. However, neither the arguments given within this effective QCD-related model in Ref. [32], nor the arguments presented within QCD itself in the work of Ref. [29], were not heard and the ideas of the applications of the PMC approximants without using $\{\beta\}$ expansion of the photon anomalous dimension continue to be applied to the associated Adler function quantities (see, e.g., Refs. [40–43] and the citations in the related discussions in even experimentally related works).

To our understanding, the opinion of the authors of, e.g., Refs. [23,31,40–43] may be summarized in the form of the following statements: Since the anomalous dimension $\gamma(a_s)$ is scheme invariant in the class of the MS-like schemes, it is associated with the renormalization of the QED coupling by the QCD corrections only. These corrections are not related to the running of the strong coupling constant and thus the photon anomalous dimension RG function should be considered as a conformal contribution during the PMC scale setting analysis.

Let us try to clarify once more unheard arguments given previously in Refs. [29,32,39] by rephrasing them in other way as follows:

(i) We disagree that $\gamma(a_s)$ is not associated with the renormalization of the strong coupling constant. On the contrary, the QCD photon anomalous dimension is inseparably related with the renormalization of the QCD charge. Indeed, the coefficients of $\gamma(a_s)$ are expressed through the first-order pole terms $Z_{n+1,-1}$ of $Z(a_s)$ (4),

$$\gamma(a_s) = -\varepsilon Z(a_s) + \frac{dZ(a_s)}{d\ln\mu^2}$$
$$= -\varepsilon Z(a_s) + (-\varepsilon a_s + \beta(a_s))\frac{\partial Z(a_s)}{\partial a_s}$$
$$= -\sum_{n\geq 0}(n+1)Z_{n+1,-1}a_s^n. \quad (27)$$

This fact directly follows from the condition of cancellation of the divergent terms in Eq. (3) and from the relation between the bare coupling $a_{sB}$ and the renormalized one $a_s$.

However, the renormalization prescription (3) allows one to present the PT coefficients of the anomalous dimension $\gamma(a_s)$ through the terms $\Pi_{p,k}$ of the $\varepsilon$ expansion of the bare polarization operator $\Pi_B(L, a_{sB})$ (5) as well,

$$\gamma_0 = \Pi_{1,-1}, \quad (28a)$$

$$\gamma_1 = 2\Pi_{2,-1}, \quad (28b)$$

$$\gamma_2 = 3\Pi_{3,-1} - 3\beta_0\Pi_{2,0}, \quad (28c)$$

$$\gamma_3 = 4\Pi_{4,-1} - 8\beta_0\Pi_{3,0} - 2\beta_1\Pi_{2,0} + 4\beta_0^2\Pi_{2,1}, \quad (28d)$$

$$\gamma_4 = 5\Pi_{5,-1} - 15\beta_0\Pi_{4,0} - 5\beta_1\Pi_{3,0} - \frac{5}{3}\beta_2\Pi_{2,0}$$
$$+ \frac{35}{6}\beta_0\beta_1\Pi_{2,1} + 15\beta_0^2\Pi_{3,1} - 5\beta_0^3\Pi_{2,2}. \quad (28e)$$

Here the total prefactor $d_R\sum_f Q_f^2$ on the rhs of Eqs. (28a)–(28e) is omitted. The expressions (28a)–(28e) reveal explicitly the structure of the $\{\beta\}$ expansion in powers of the coefficients of the QCD RG $\beta$ function. Moreover, the $\beta$-dependent terms in Eqs. (28c)–(28e) appear after the QCD charge renormalization only. Therefore, the total expression for the photon anomalous dimension should not be considered as the conformal part of $D(Q^2)$ function. This statement is clarified below.

(ii) Since the strong coupling is running, then the QCD photon anomalous dimension is not a scale-invariant object. Indeed, $d\gamma(a_s)/d\ln(Q^2) = \beta(a_s)\partial\gamma(a_s)/\partial a_s = -\beta_0\gamma_1 a_s^2 - (2\beta_0\gamma_2 + \beta_1\gamma_1)a_s^3 + \ldots \neq 0$, where $a_s = a_s(Q^2)$. Therefore, the scheme-invariance of its *coefficients* in the class of the gauge-invariant renormalization MS-like schemes [due to the relations $\gamma_k = -(k+1)Z_{k+1,-1}$ (27) and the scheme invariance of the first-order pole terms in $Z(a_s)$] is not the argument against $\{\beta\}$ decomposition of $\gamma(a_s)$.

Let us repeat now another serious extra argument given in Ref. [29] for applying $\beta$ expansion to $\gamma(a_s)$.

(iii) In the QED limit, the term $\tilde{d}_2[0]$ (B2b) becomes equal to $\tilde{d}_2^{\text{QED}}[0] = -3/32 - 11/48N$, where $N$ is the number of the charged leptons.

This expression is $N$ dependent and does not correspond to Rosner's result [63] of calculating of the divergent part of the photon field renormalization constant $Z_{ph}$ in the quenched QED, formulated in the diagrammatic level in Ref. [64]. In this finite approximation, the constant $Z_{ph}$ does not contain the internal subgraphs renormalizing electromagnetic charge. The result of this work is $(Z_{ph}^{-1})_{\text{div}} = \frac{a_B}{3}(1 + \frac{3}{4}a_B \boxed{-\frac{3}{32}} a_B^2) \ln \frac{M^2}{m^2}$, where $a_B = \alpha_B/\pi$, $\alpha_B$ is the bare fine-structure constant, $m$ is the lepton mass, and $M$ is the large-scale cutoff mass. The boxed term does not match the expression for $\tilde{d}_2^{\text{QED}}[0]$, obtained when the photon anomalous dimension is not $\{\beta\}$ decomposed, but it is in full agreement with the result for $d_2^{\text{QED}}[0]$, following from the $U(1)$ limit of Eq. (23a) for $\gamma_2[0]$ at $C_F = 1$ and $C_A = 0$.

There is also the $\mathcal{N} = 1$ SUSY QCD argument in favor of the title of this section:

(iv) The NSVZ-like relation for the Adler function in the $\mathcal{N} = 1$ SUSY QCD derived in Ref. [65] and its detailed consideration at the three-loop level made in Ref. [66] serve as the extra arguments in favor of the $\{\beta\}$ expansion of the SUSY analog of the photon anomalous dimension, namely, the anomalous dimension of the matter superfields. Indeed, the Novikov-Shifman-Vainshtein-Zakharov relation will be violated at the three-loop level if one does not decompose this anomalous dimension in the first coefficient of the corresponding $\beta$ function.[3]

Note that, if we consider the Adler function defined in Eq. (15) directly, without involving Eq. (13) linking the functions $D(a_s)$, $\gamma(a_s)$, and $\Pi(a_s)$, then $\{\beta\}$ expansion for its PT expression should not depend on $\gamma(a_s)$ and $\Pi(a_s)$. Moreover, from a formal point of view, for the $\{\beta\}$ decomposition there is no principal difference, for example, between the PT series for the $D(Q^2)$ function, the Bjorken polarized sum rule, or the static interaction potential of the heavy quark-antiquark pair.

However, the results of the $\{\beta\}$ expansion for the Adler function, presented in Ref. [31], depend on $\gamma(a_s)$ in any case. At the same time, in the same paper, the $\{\beta\}$ decomposition for the static QCD Coulomb-like potential, calculated analytically at the three-loop level in Ref. [67], is implemented on general grounds as in Ref. [30] as well.

Therefore, the agreement of the results of $\{\beta\}$ expansion for the static potential derived in Ref. [31] with those obtained in Ref. [30] in the framework of our formalism, is rather natural.

Thus, the photon anomalous dimension is the convenient ingredient for analytical calculations, but it should not affect the structure of the $\{\beta\}$ expansion of the Adler function.

If one follows the logic of works [23,31,40–43] and does not decompose the quantity $\gamma(a_s)$ in powers of $\beta_k$ coefficients, then analytical expressions for the analogous to $d_k[\ldots]$ coefficients will be different. We denote these terms by $\tilde{d}_k[\ldots]$ to distinguish them from ours $d_k[\ldots]$. For comparison of their analytical structure, see expressions for $d_k[\ldots]$ and $\tilde{d}_k[\ldots]$ in Appendix B.

Note that the explicit analytical expressions for $\gamma_4$ and $\Pi_3$ contain the Riemann $\zeta_4$ contributions [58], which, however, are mutually canceled out in $d_4$ [57], i.e.,

$$d_4^{(\zeta_4)} = \gamma_4^{(\zeta_4)} + 3\beta_0 \Pi_3^{(\zeta_4)} = 0. \qquad (29)$$

If we properly expand $\gamma_4$ and $\Pi_3$ [in accordance with Eqs. (21d) and (24)], we will naturally arrive at the absence of the $\zeta_4$ contributions in the expression for $d_4[0]$ [28,29]. However, if one assumes that $\gamma_4$ is a scale-invariant term, then the coefficient $\tilde{d}_4[0]$ [see Eq. (B2g)] will definitely contain $\zeta_4$ contributions. This fact contradicts the consequences of the no-$\pi$ theorem [68], explaining why the $\zeta_4$ contribution should appear in the expressions for higher-order PT corrections to the Adler function starting from the coefficient $d_5$ only.

Let us now discuss the consequences stemming from the results of Refs. [23,31,40–42] for the terms $\tilde{d}_k[\ldots]$ obtained when $\gamma(a_s)$ is not $\{\beta\}$ expanded.

(v) In this case, the $\{\beta\}$ decomposition of the coefficients $d_k$ of the Adler function has the following form:

$$d_2 = \beta_0 \tilde{d}_2[1] + \tilde{d}_2[0], \qquad (30a)$$

$$d_3 = \underbrace{\beta_0^2 \tilde{d}_3[2]}_{=0} + \beta_1 \tilde{d}_2[1] + 2\beta_0 \tilde{d}_3[1] + \tilde{d}_3[0], \qquad (30b)$$

$$d_4 = \underbrace{\beta_0^3 \tilde{d}_4[3]}_{=0} + \underbrace{3\beta_0^2 \tilde{d}_4[2]}_{=0} + 3\beta_0 \tilde{d}_4[1] + \underbrace{\frac{5}{2}\beta_1 \beta_0 \tilde{d}_3[2]}_{=0}$$
$$+ 2\beta_1 \tilde{d}_3[1] + \beta_2 \tilde{d}_2[1] + \tilde{d}_4[0], \qquad (30c)$$

where curly brackets indicate terms with identically zero coefficients. It should be emphasized that this representation does not correspond to the well-known renormalon asymptotics $d_{k+1} \sim \beta_0^k k!$ at $k \gg 1$ for higher-order PT coefficients of the Adler function in the large-$\beta_0$ approximation (see, e.g., [1,2,52]). Indeed, all terms $\tilde{d}_{k+1}[k]$ in Eqs. (30b) and (30c) at $k \geq 2$ are *identically* equal to zero. Thus, if we do not decompose the coefficients $\gamma_k$ and $\Pi_k$ in powers of the RG $\beta$-function coefficients, then we will not reproduce the large-$\beta_0$ asymptotics for $d_k$ in any order of PT starting from $k = 3$.

---

[3] It may be interesting to get the arguments in favor of this statement in the $\mathcal{N} = 1$ SUSY QCD at the four-loop level.

The leading renormalon chain contribution, whose explicit general formula for arbitrary order $k$ follows from analytical results, given in Refs. [7,62], is fixed correctly when the photon anomalous dimension undergoes the $\{\beta\}$-expansion procedure only. In its turn, in Refs. [23,31,40–42] the missing $n_f$-dependent contributions are hidden in the expressions for nonzero terms $\tilde{d}_2[0], \tilde{d}_2[1]; \tilde{d}_3[0], \tilde{d}_3[1]; \tilde{d}_4[0], \tilde{d}_4[1]$ in Eqs. (30a)–(30c).

One more important point, which follows from totally decomposed $\beta$-expanded representation for the Adler function, is the recovery of its original BLM prescription NLO expression. Thus, the worries of Ref. [64] on nonrecovery of the BLM results within the application of the $\mathcal{R}_\delta$ procedure to the PT expression for the Adler function without proper expansion of the PT QCD series for the photon anomalous dimension $\gamma(a_s)$ considered in Ref. [23], critically commented in the more detailed PMC-related work of Ref. [31], turned out to have rather solid background. Since after applying the multiple $\beta$-function representation to the Adler function formulated in Eq. (19), we reproduce its NLO BLM prescription expression, we will call this the PMC/BLM approach.

## V. APPLICATION OF THE PMC/BLM APPROXIMANTS TO THE ADLER FUNCTION

### A. Modified PMC expressions

At the first stage of the BLM presription application, one should consider the scale transformations $\mu \to \mu'$ and introduce the shift parameter $\Delta = L - L' = \ln(\mu^2/\mu'^2)$, where $L' = \ln(\mu'^2/Q^2)$.

Using now the scaling operator (which may also be called the dilatation operator), one can obtain the relation between $a_s(\mu^2)$ and $a_s(\mu'^2)$ in the following form, considered previously in Ref. [4,32]:

$$a_s(\mu^2) = a_s(\exp(\Delta) \cdot \mu'^2) = \exp\left(\Delta \frac{d}{d\ln\mu'^2}\right) a'_s$$

$$= \exp\left(\Delta \beta(a'_s) \frac{\partial}{\partial a'_s}\right) a'_s$$

$$= a'_s + \frac{\Delta}{1!}\beta(a'_s) + \frac{\Delta^2}{2!}\beta(a'_s)\frac{\partial}{\partial a'_s}\beta(a'_s)$$

$$+ \frac{\Delta^3}{3!}\beta(a'_s)\frac{\partial}{\partial a'_s}\left(\beta(a'_s)\frac{\partial}{\partial a'_s}\beta(a'_s)\right) + \ldots, \quad (31)$$

where $a'_s = a_s(\mu'^2)$.

At the next step, we choose the PMC/BLM scale shift $\Delta$ as a PT series in powers of $\beta_0 a'_s$,

$$\Delta = \ln\left(\frac{\mu^2}{\mu'^2}\right) = \Delta_0 + \sum_{k\geq 1}\Delta_k(\beta_0 a'_s)^k. \quad (32)$$

Taking into account this representation, one can rewrite the relation (31) in the fourth order of approximation in the following form:

$$a_s = a'_s - \beta_0 \Delta_0 a'^2_s + (\beta_0^2 \Delta_0^2 - \beta_1 \Delta_0 - \beta_0^2 \Delta_1) a'^3_s$$

$$+ \left(\frac{5}{2}\beta_0\beta_1 \Delta_0^2 - \beta_0\beta_1\Delta_1 + 2\beta_0^3 \Delta_0 \Delta_1\right.$$

$$\left. - \beta_0^3 \Delta_0^3 - \beta_2 \Delta_0 - \beta_0^3 \Delta_2\right) a'^4_s. \quad (33)$$

Using now Eq. (33) at $\mu'^2 = Q^2$, bearing in mind the RG invariance of the Adler function and its $\{\beta\}$-expansion pattern (18a)–(18f), it is possible to get the expressions for the coefficients $d'_k$ of the $D(a'_s)$ function, normalized at the new scale, in the form given in Refs. [4,32],

$$d'_1 = d_1[0], \tag{34a}$$

$$d'_2 = \beta_0(d_2[1] - \Delta_0 d_1[0]) + d_2[0], \tag{34b}$$

$$d'_3 + \delta_f(d_3^{\rm SI})' = \beta_0^2(d_3[2] - 2\Delta_0 d_2[1] + \Delta_0^2 d_1[0] - \Delta_1 d_1[0]) + \beta_1(d_3[0,1] - \Delta_0 d_1[0]) + \beta_0(d_3[1] - 2\Delta_0 d_2[0]) + d_3[0]$$
$$+ \delta_f d_3^{\rm SI}[0], \tag{34c}$$

$$d'_4 + \delta_f(d_4^{\rm SI})' = \beta_0^3(d_4[3] - 3\Delta_0 d_3[2] + 3\Delta_0^2 d_2[1] - 2\Delta_1 d_2[1] + 2\Delta_0\Delta_1 d_1[0] - \Delta_0^3 d_1[0] - \Delta_2 d_1[0])$$
$$+ \beta_0\beta_1(d_4[1,1] - 3\Delta_0 d_3[0,1] - 2\Delta_0 d_2[1] + 5\Delta_0^2 d_1[0]/2 - \Delta_1 d_1[0]) + \beta_2(d_4[0,0,1] - \Delta_0 d_1[0])$$
$$+ \beta_0^2(d_4[2] - 3\Delta_0 d_3[1] + 3\Delta_0^2 d_2[0] - 2\Delta_1 d_2[0]) + \beta_1(d_4[0,1] - 2\Delta_0 d_2[0]) + \beta_0(d_4[1] - 3\Delta_0 d_3[0])$$
$$+ \beta_0 \delta_f(d_4^{\rm SI}[1] - 3\Delta_0 d_3^{\rm SI}[0]) + d_4[0] + \delta_f d_4^{\rm SI}[0], \tag{34d}$$

where $\delta_f = (\sum Q_f)^2 / \sum Q_f^2$. Recall that the coefficients $d_k[\ldots]$ are presented in Appendix B.

Setting initially

$$\Delta_0 = \Delta_{\text{BLM}} = \frac{d_2[1]}{d_1[0]} = \left(\frac{33}{8} - 3\zeta_3\right) C_F, \quad (35)$$

one can introduce the energy scale $Q_0^2$,

$$Q_0^2 = Q^2 \exp(-\Delta_0). \quad (36)$$

At this new scale, the expression for the Adler function reads

$$D(Q^2) = 3\sum_f Q_f^2 (1 + d_1[0]a_s(Q_0^2) + d_2[0]a_s^2(Q_0^2) + \mathcal{O}(a_s^3(Q_0^2))). \quad (37)$$

Further on, taking into account relation (35) and absorbing the remaining $\beta_i$-dependent contributions in Eq. (34c) into parameter $\Delta_1$, we arrive at the following expression:

$$\beta_0 \Delta_1(n_f) = \beta_0 \left(\frac{d_3[2]}{d_1[0]} - \frac{d_2^2[1]}{d_1^2[0]}\right) + \frac{d_3[1]}{d_1[0]} - \frac{2d_2[0]d_2[1]}{d_1^2[0]} + \frac{\beta_1}{\beta_0} \frac{d_3[0,1] - d_2[1]}{d_1[0]}. \quad (38)$$

Application of the PMC/BLM approach at the $\mathcal{O}(a_s^3)$ level eventually yields

$$D(Q^2) = 3\sum_f Q_f^2 (1 + d_1[0]a_s(Q_1^2) + d_2[0]a_s^2(Q_1^2) + (d_3[0] + \delta_f d_3^{\text{SI}}[0])a_s^3(Q_1^2) + \mathcal{O}(a_s^4(Q_1^2))), \quad (39)$$

where $Q_1^2$ is defined in accordance with Eqs. (32), (35), (36), and (38) as

$$Q_1^2 = Q^2 \exp(-\Delta_0 - \beta_0 \Delta_1(n_f) a_s(Q_0^2)). \quad (40)$$

Following this logic and using Eq. (34d), one can fix the parameter $\beta_0^2 \Delta_2$ as

$$\beta_0^2 \Delta_2(n_f) = \beta_0^2 \left(\frac{d_4[3]}{d_1[0]} - 3\frac{d_2[1]d_3[2]}{d_1^2[0]} + 2\frac{d_2^3[1]}{d_1^3[0]}\right) + \beta_1 \left(\frac{d_4[1,1]}{d_1[0]} - 3\frac{d_2[1]d_3[0,1]}{d_1^2[0]} + \frac{3}{2}\frac{d_2^2[1]}{d_1^2[0]} - \frac{d_3[2]}{d_1[0]}\right)$$
$$+ \beta_0 \left(\frac{d_4[2]}{d_1[0]} - 3\frac{d_3[1]d_2[1]}{d_1^2[0]} + 5\frac{d_2[0]d_2^2[1]}{d_1^3[0]} - 2\frac{d_2[0]d_3[2]}{d_1^2[0]}\right) + \frac{d_4[1]}{d_1[0]} - 3\frac{d_3[0]d_2[1]}{d_1^2[0]} + \delta_f \left(\frac{d_4^{\text{SI}}[1]}{d_1[0]} - 3\frac{d_3^{\text{SI}}[0]d_2[1]}{d_1^2[0]}\right)$$
$$- 2\frac{d_2[0]d_3[1]}{d_1^2[0]} + 4\frac{d_2^2[0]d_2[1]}{d_1^3[0]} + \frac{\beta_1^2}{\beta_0^2}\frac{d_2[1] - d_3[0,1]}{d_1[0]} + \frac{\beta_1}{\beta_0}\left(\frac{d_4[0,1] - d_3[1]}{d_1[0]} - \frac{2d_2[0]}{d_1^2[0]}(d_3[0,1] - d_2[1])\right)$$
$$+ \frac{\beta_2}{\beta_0}\frac{d_4[0,0,1] - d_2[1]}{d_1[0]}. \quad (41)$$

In this case, instead of the expressions (39) and (40), we obtain their higher-order counterparts,

$$D(Q^2) = 3\sum_f Q_f^2 (1 + d_1[0]a_s(Q_2^2) + d_2[0]a_s^2(Q_2^2) + (d_3[0] + \delta_f d_3^{SI}[0])a_s^3(Q_2^2) + (d_4[0] + \delta_f d_4^{SI}[0])a_s^4(Q_2^2) + \mathcal{O}(a_s^5(Q_2^2))), \quad (42)$$

$$Q_2^2 = Q^2 \exp(-\Delta_0 - \beta_0 \Delta_1(n_f) a_s(Q_1^2) - \beta_0^2 \Delta_2(n_f) a_s^2(Q_1^2)). \quad (43)$$

In a particular case of the $SU(3)$ color gauge group, the numerical forms of the parameters $\Delta_0$, $\beta_0 \Delta_1$, and $\beta_0^2 \Delta_2$ are defined correspondingly and are included into determination of the scale $Q_2^2$ in Eq. (43), are defined as

$$\Delta_0 = \frac{11}{2} - 4\zeta_3 \approx 0.6918, \quad (44a)$$

$$\beta_0 \Delta_1(n_f) = \beta_0 \left(\frac{119}{36} + \frac{56}{3}\zeta_3 - 16\zeta_3^2\right) + \frac{51}{8} - \frac{47}{3}\zeta_3 + \frac{50}{3}\zeta_5$$
$$\approx 2.6249\beta_0 + 4.8249, \quad (44b)$$

$$\beta_0^2 \Delta_2(n_f) = -3.599\beta_0^2 + 2.386\beta_1 + 7.128\beta_0 - 54.535 - 0.292\delta_f. \quad (44c)$$

Note that, in reality, the expressions (44b) and (44c) do not contain the terms proportional to the factor $\beta_1/\beta_0$ and $\beta_1^2/\beta_0^2$, $\beta_1/\beta_0$, $\beta_2/\beta_0$, which are included in their corresponding analytical forms, presented in Eqs. (38) and (41). This pleasant fact is the consequence of relationships (20a) and (20b) stemming from the multiple $\beta$-function expansion (19), proposed in Ref. [28].

Using the values of the coefficients $d_k[\ldots]$, given in Appendix B and originally obtained in Ref. [28] within the same decomposition (19) of the Adler function in powers of $\beta(a_s)/a_s$, advocated in our work, we get its numerical expression in the case of the $SU(3)$ color group relevant for physical QCD,

$$D(Q^2) = 3\sum_f Q_f^2 \bigg(1 + a_s(Q_2^2) + \frac{1}{12}a_s^2(Q_2^2) + (-23.2227 - 0.4132\delta_f)a_s^3(Q_2^2)$$
$$+ (81.1571 + 0.0802n_f - 2.7804\delta_f)a_s^4(Q_2^2) + \mathcal{O}(a_s^5(Q_2^2))\bigg). \tag{45}$$

It is worth mentioning that the numerical results of Eq. (45) were previously presented in Ref. [29].

The magnitudes of their $\mathcal{O}(a_s^2)$ and $\mathcal{O}(a_s^3)$ coefficients are in agreement with the ones received in Ref. [14] with help of the generalized BLM prescription and the large-$n_f$ expansion [see the related work of Ref. [16] where the numerical expression for the related $\mathcal{O}(a_s^4)$ coefficient in Eq. (45) was found].

Our expression (45) should be also compared with its counterpart following from the PMC-type considerations of Refs. [23,31,40–43] with the $\{\beta\}$-nonexpanded photon anomalous dimension,

$$D(Q^2) = 3\sum_f Q_f^2(1 + a_s(\tilde{Q}_2^2) + (2.6042 - 0.1528n_f)a_s^2(\tilde{Q}_2^2) + (9.7418 - 2.0426n_f - 0.0198n_f^2 - 0.4132\delta_f)a_s^3(\tilde{Q}_2^2)$$
$$+ (41.0141 - 12.9110n_f + 0.4887n_f^2 + 0.0045n_f^3 + (-2.3829 - 0.0241n_f)\delta_f)a_s^4(\tilde{Q}_2^2) + \mathcal{O}(a_s^5(\tilde{Q}_2^2))), \tag{46}$$

where we do not specify the explicit form of the corresponding scale $\tilde{Q}_2^2$, which does not coincide with $Q_2^2$, is defined in a similar way as indicated in Eq. (43).

As we have already discussed above, this coefficients in the expression (46) are $n_f$-dependent ones. This essential difference of Eq. (46) from Eq. (45) is the consequence of the not applied $\{\beta\}$-expansion procedure to the photon anomalous dimension $\gamma(a_s)$ in Refs. [23,31,40–43]. This fact was critically commented on in Sec. IV of this work.

We also present here the numerical $\overline{\text{MS}}$-scheme result for the Adler function, which follows from the analytical $\mathcal{O}(a_s^4)$ expression, obtained in [57,58] and confirmed in [69]

$$D(Q^2) = 3\sum_f Q_f^2(1 + a_s(Q^2) + (1.9857 - 0.1153n_f)a_s^2(Q^2) + (18.2427 - 4.2158n_f + 0.0862n_f^2 - 0.4132\delta_f)a_s^3(Q^2)$$
$$+ (135.7916 - 34.4402n_f + 1.8753n_f^2 - 0.0101n_f^3 + (-5.9422 + 0.1916n_f)\delta_f)a_s^4(Q^2) + \mathcal{O}(a_s^5(Q_2^2))). \tag{47}$$

It is worth clarifying that the leading large $n_f$ contributions to Eq. (47) do agree with the numerical form of the analytical QED result, obtained previously in Ref. [62], but disagree with the analogous numbers, given in Eq. (46) above. This is a consequence of the fact that although the nonexpanded expression for $\gamma(a_s)$ contains a substantial part of the renormalon-related contributions to the Adler function, they are contained in the expression for the vacuum polarization function as well. The latter are absorbed into the scale $\tilde{Q}_2^2$ while the PMC procedure is applied and the renormalon contributions to $\gamma(a_s)$ remain. From our point of view, such a variant of the realization of the PMC approach, generally representing interesting and important ideas, is not fully theoretically justified.

### B. Energy dependence of the PMC/BLM and $\overline{\text{MS}}$-scheme Adler function approximants

#### 1. PMC/BLM inputs

Let us now specify what we mean under the expressions for the expansion parameters $a_s(Q_0^2)$, $a_s(Q_1^2)$, and $a_s(Q_2^2)$ in the NLO, N²LO, and N³LO PMC/BLM Adler function approximants, which are presented in Eqs. (37), (39), and (42) above. They correspond to the truncated at definite orders of PT inverse log representation of the $\overline{\text{MS}}$-scheme QCD coupling constant taken, e.g., from Eq. (9.5) of the QCD PDG review of Ref. [70] with the NLO, N²LO, and N³LO energy scales $Q_0^2$, $Q_1^2$, and $Q_2^2$ being fixed at the related orders of these representations through the relatively applied PMC/BLM expressions of Eqs. (36), (40), and (43).

In concrete applications, these ways of fixation can be rewritten through the unique $\overline{\text{MS}}$-scheme representation of the QCD coupling constant related to the arbitrary energy scale $Q^2$, but with the appropriately redefined expressions of the $\overline{\text{MS}}$-scheme scale parameter, namely,

$$\Lambda_{\text{NLO}}^{(\text{BLM})}(n_f) = \Lambda_{\text{NLO}}^{(n_f)} \cdot \exp\left[-\frac{1}{2}\Delta_0\right], \tag{48a}$$

$$\Lambda_{\text{N}^2\text{LO}}^{(\text{PMC})}(n_f) = \Lambda_{\text{N}^2\text{LO}}^{(n_f)} \cdot \exp\left[-\frac{1}{2}\left(\Delta_0 + \beta_0\Delta_1(n_f)a_s^{\text{NLO}}(Q^2/(\Lambda_{\text{NLO}}^{(\text{BLM})})^2)\right)\right], \tag{48b}$$

$$\Lambda_{\text{N}^3\text{LO}}^{(\text{PMC})}(n_f) = \Lambda_{\text{N}^3\text{LO}}^{(n_f)} \cdot \exp\left[-\frac{1}{2}\left(\Delta_0 + \beta_0\Delta_1(n_f)a_s^{\text{N}^2\text{LO}}(Q^2/(\Lambda_{\text{N}^2\text{LO}}^{(\text{PMC})})^2) + \beta_0^2\Delta_2(n_f)\left(a_s^{\text{N}^2\text{LO}}(Q^2/(\Lambda_{\text{N}^2\text{LO}}^{(\text{PMC})})^2)\right)^2\right)\right], \tag{48c}$$

where $\Lambda_{\text{NLO}}^{(n_f)}$, $\Lambda_{\text{N}^2\text{LO}}^{(n_f)}$, $\Lambda_{\text{N}^3\text{LO}}^{(n_f)}$ are the expressions for the QCD scale parameter defined in the $\overline{\text{MS}}$ scheme in the corresponding order of PT, while $\Delta_0$, $\beta_0\Delta_1(n_f)$, $\beta_0^2\Delta_2(n_f)$ are defined by Eqs. (44a)–(44c) presented above.

We will use the appropriately truncated RG-improved expressions for the running QCD coupling $a_s(Q^2)$ through the $\ln(Q^2/\Lambda^{(n_f)2})$ terms and the coefficients of the QCD $\beta$ function, taking into account its N$^3$LO four-loop coefficient, analytically evaluated in Ref. [71] and confirmed in Ref. [72]. In fact, at present, the five-loop $\beta_4$ term of the corresponding $\beta$ function is also known. It was analytically evaluated in Ref. [73] and confirmed in Refs. [74,75]. However, for the sake of consistency with the orders of truncation of the PT approximations for the Adler function, this term will be not taken into account.

### 2. The $\overline{\text{MS}}$-scheme benchmarks

We will fix as the initial normalization point the $\tau$-lepton pole mass $M_\tau = 1776.8$ MeV, will consider $n_f = 3$ number of active flavors, and will use the rounded strong coupling constant value $\alpha_s(M_\tau^2) = 0.312$, extracted in Ref. [76] from the QCD sum rules analysis of the ALEPH Collaboration $\tau$-lepton decay data. In view of the qualitative aims of our studies to be presented below, we will neglect the consideration of the effects of theoretical and experimental uncertainties. We note, however, that the result of Ref. [76] falls into the uncertainty bands of the related results, independently obtained in Ref. [77] from a more detailed reanalysis of the same ALEPH data.

Considering now the properly truncated at the NLO, N$^2$LO, and N$^3$LO representations $\alpha_s(M_\tau^2)$ through the inverse powers of logarithms from $M_\tau^2/\Lambda^{(3)2}$ ratio, we arrive at the following, of course rather rough, values for the $\overline{\text{MS}}$ scale QCD parameter $\Lambda^{(3)}$ at $n_f = 3$:

$$\Lambda_{\text{NLO}}^{(3)} = 361 \text{ MeV}, \quad \Lambda_{\text{N}^2\text{LO}}^{(3)} = 330 \text{ MeV}, \quad \Lambda_{\text{N}^3\text{LO}}^{(3)} = 325 \text{ MeV}.$$

To transform them to the cases of $n_f = 4$ and $n_f = 5$ effective number of quark flavors, we will use the expressions for the threshold transformation formulas available from the results of Refs. [78–81] with the corresponding matching scales fixed at $\sqrt{Q^2} = 2\bar{m}_c(\bar{m}_c^2) = 2.54$ and $\sqrt{Q^2} = 2\bar{m}_b(\bar{m}_b^2) = 8.36$ GeV. They are related to the following values of the $\overline{\text{MS}}$ scheme running $c$- and $b$-quark masses, $\bar{m}_c(\bar{m}_c^2) = 1.27$ and $\bar{m}_b(\bar{m}_b^2) = 4.18$ GeV, taken from the PDG (2022) volume of Ref. [70].

Following these steps, we obtain the related cases of $n_f = 4$ and $n_f = 5$ numbers of active flavor sets of the numerical values of the $\overline{\text{MS}}$-scheme scale parameter,

$$\Lambda_{\text{NLO}}^{(4)} = 315, \quad \Lambda_{\text{N}^2\text{LO}}^{(4)} = 286, \quad \Lambda_{\text{N}^3\text{LO}}^{(4)} = 282 \text{ MeV}$$

and

$$\Lambda_{\text{NLO}}^{(5)} = 223, \quad \Lambda_{\text{N}^2\text{LO}}^{(5)} = 205, \quad \Lambda_{\text{N}^3\text{LO}}^{(5)} = 203 \text{ MeV}.$$

The choice of the concrete threshold energies is, of course, ambiguous and will introduce additional inaccuracies [82]. However, these effects are also not substantial for our aims and we will neglect them as well in our considerations.

Using the given above values of the $\overline{\text{MS}}$-related QCD scale, the expressions from Eqs. (48a)–(48c), the inverse logarithmic representation of strong coupling in the NLO, N$^2$LO, and N$^3$LO approximations, and the given in Eqs. (47) and (45) explicit expressions for $D(Q^2)$ in the $\overline{\text{MS}}$ scheme and within the PMC/BLM procedure, we can get the corresponding energy dependence of the Adler function for the $\overline{\text{MS}}$ and PMC/BLM approximants to compare them with each other.

It was also checked that the evolution of the taken value $\alpha_s(M_\tau^2) = 0.312$ up to the mass $M_Z = 91.188$ GeV [70] of $Z^0$ boson in QCD at the $\mathcal{O}(\alpha_s^4)$ level yields $\alpha_s(M_Z^2) = 0.1175$. It is consistent with the results of [76] and with the average value of PDG (2022) [70] with unaccounted by us uncertainties. Thus, we hope that we presented enough arguments for convincing careful readers that our qualitatively aimed study is quantitatively self-consistent.

### 3. The phenomenologically relevant $n_f = 3$ case

To illustrate the characteristic behavior of the Adler function approximants in the $\overline{\text{MS}}$ scheme and in the PMC/BLM approach in the case of $n_f = 3$, we consider the region of the Euclidean transferred momentum $1.5 \leq \sqrt{Q^2} \leq 2.4$ GeV, where the lower-energy scale is slightly smaller than the $\tau$-lepton mass and the upper one is a bit smaller than twice the charm-quark mass.

Note that, in the Minkowskian timelike domain in the similar energy region $1.84 \leq \sqrt{s} \leq 3.88$ GeV, the subprocess of the production of the light quark-antiquark $u, d, s$ pairs dominates in the $e^+e^-$ annihilation into hadrons process. In this domain, the experimental data for the total cross section of the discussed subprocess were extracted from measurements provided by KEDR [83,84] and BESIII [85] Collaborations.

Taking into account the results of studies and benchmarks presented above, we obtain Fig. 1(a), demonstrating the energy behavior of the NLO, N²LO, and N³LO massless approximants for the Adler function in the $\overline{\text{MS}}$ scheme (47) and in the PMC/BLM approach. For comparison, the Born quark-parton result $3 \sum Q_f^2$ is presented there as well.

Let us comment on definite consequences following from the comparison of the behavior of various curves presented in Fig. 1(a).

(1) One can see that the NLO PT corrections to the Adler function are leading to the corrections, which are quantitatively defining main contributions in both $\overline{\text{MS}}$ and PMC/BLM cases.

(2) While taking into account higher-order PT corrections, we observe the characteristic difference in the fine structure of sets of the $\overline{\text{MS}}$ scheme and PMC/BLM approximants. Indeed, the $\overline{\text{MS}}$ results satisfy the inequalities $D_{\text{Born}} < D_{\text{NLO}}(Q^2) < D_{\text{N}^2\text{LO}}(Q^2) < D_{\text{N}^3\text{LO}}(Q^2)$, whereas for PMC/BLM we have $D_{\text{Born}} < D_{\text{NLO}}(Q^2)$, but $D_{\text{NLO}}(Q^2) > D_{\text{N}^2\text{LO}}(Q^2)$ and $D_{\text{N}^2\text{LO}}(Q^2) < D_{\text{N}^3\text{LO}}(Q^2)$.

(3) It is interesting that the sign structure of the related N³LO PT QCD expressions changes from the $++++$ pattern in the $\overline{\text{MS}}$-scheme case to the pattern $++-+$ in the PMC/BLM case.

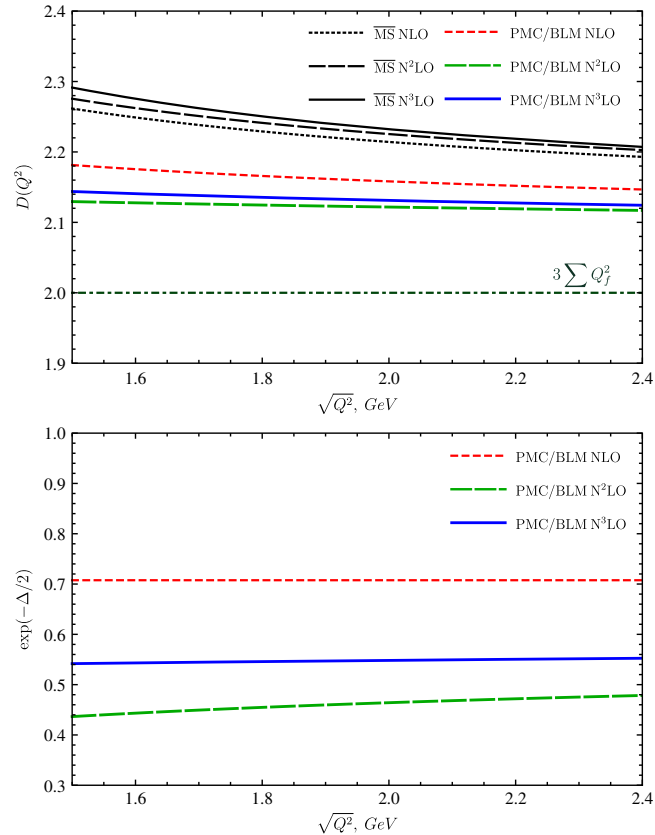

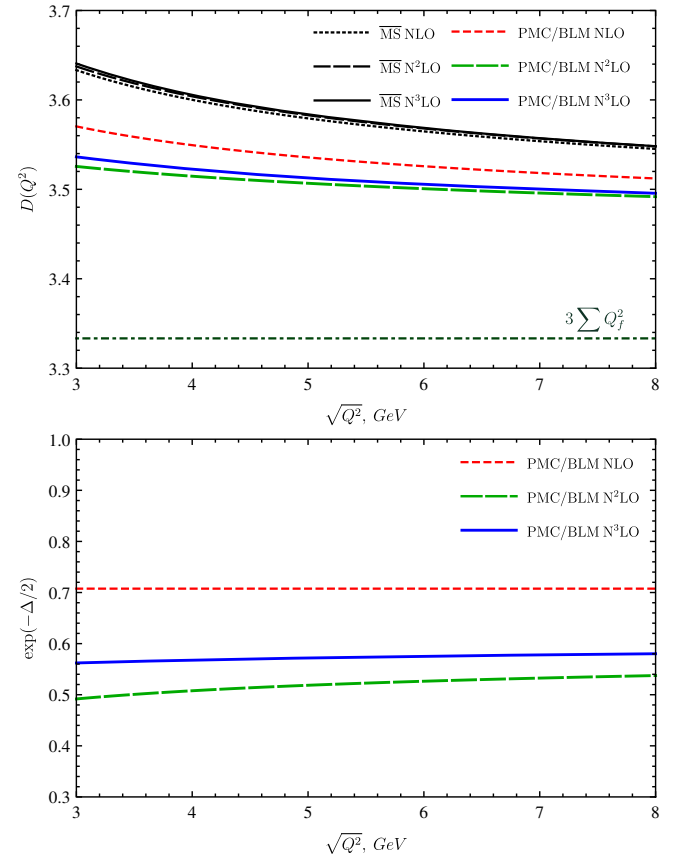

FIG. 1. (a) The dependence of the PT Adler function $D(Q^2)$ on $\sqrt{Q^2}$ at $n_f = 3$ in the massless limit. The Born factor $3 \sum Q_f^2$ is presented as well. (b) The dependence of the factor $\exp(-\Delta/2)$ on $\sqrt{Q^2}$ at $n_f = 3$.

FIG. 2. (a) The dependence of the PT Adler function $D(Q^2)$ on $\sqrt{Q^2}$ at $n_f = 4$ in the massless limit. The Born factor $3 \sum Q_f^2$ is presented as well. (b) The dependence of the factor $\exp(-\Delta/2)$ on $\sqrt{Q^2}$ at $n_f = 4$.

(4) The PMC/BLM approximants are located considerably below the $\overline{\text{MS}}$ ones and as was expected are really rather flat [see Fig. 1(b) as well].

The recently presented [86] detailed phenomenologically related analysis of the Davier-Hoecker-Malaescu-Zhang compilation [87] of the experimental data for the total cross section of the $e^+e^-$ annihilation to hadrons process (though without the published data provided by KEDR and BESIII Collaborations [84,85] and most recent very interesting CMD-3 Collaboration new data [88]), and the less detailed described analysis of Refs. [89,90] of the previous $e^+e^-$ to hadrons experimental data, demonstrate that in this region of energies the experimentally related expression for the Adler function is higher than even the depicted massless $\overline{\text{MS}}$ approximants. In Ref. [86], it is clearly demonstrated that taking into account the effects of s-quark mass-dependent corrections, c-quark mass-dependent corrections, and the nonperturbative power-suppressed corrections are minimizing the difference between the $\overline{\text{MS}}$ D-function expression and Adler D-function experimentally related behavior, extracted in Refs. [86,89,90] from the concrete experimental data of the $e^+e^-$ colliders.

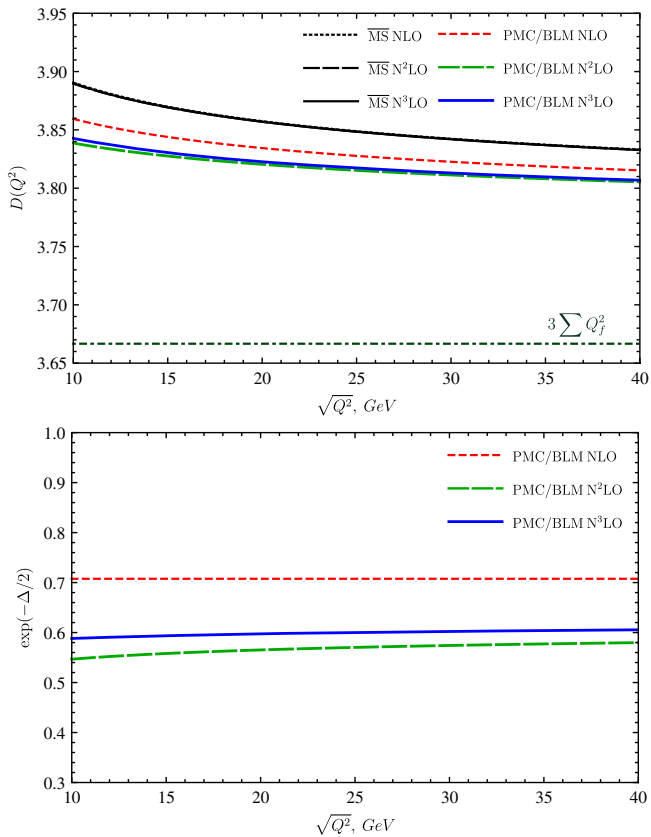

FIG. 3. (a) The dependence of the PT Adler function $D(Q^2)$ on $\sqrt{Q^2}$ at $n_f = 5$ in the massless limit. The Born factor $3\sum Q_f^2$ is presented as well. (b) The dependence of the factor $\exp(-\Delta/2)$ on $\sqrt{Q^2}$ at $n_f = 5$.

However, as seen from Fig. 1(a), the application of the PMC/BLM procedure to the massless $\overline{\text{MS}}$ Adler function PT approximations is leading to moving the considered curves lower away from the experimentally based results for the Adler function in the considered kinematical region. Therefore, we do not recommend using the PMC/BLM approach in the process of comparison with the existing experimental data and, in particular, the ones provided by the $e^+e^-$ colliders.

The previously made within the PMC approach indication on improvement of the agreement of the corresponding PMC PT QCD approximants with the available experimental data for different processes (see the related papers, starting from, e.g., Ref. [31] up to Ref. [43]) may be at least rather questionable.

Indeed, even the claims that the PMC-type QCD expressions, related to the totally non-$\beta$-expanded Adler function representation of Eq. (46), do not contain the definite parts of the effects, governed by the leading-order renormalon chain contribution, as shown in the process of this work, turned out to be not correct (compare the material presented in Appendix B).

However, it is pleasant that our realizations of the PMC/BLM ideas seem to be self-consistent. Notice the flat behavior of the PMC/BLM approximants and of the related scale transformation factors, depicted at Figs. 1(a) and 1(b). The latter one demonstrates almost independence on the transferred momentum of the PMC/BLM exponential factor $\exp(-\Delta/2)$.

These facts are related to the scale invariance of the coefficients $d_k[0]$, obtained within the $\{\beta\}$ expansion (19), and to the absorbtion of the coefficients proportional to the QCD $\beta$-function terms into the related PMC/BLM scales.

### C. Theoretically useful $n_f = 4, 5$ cases

Similar to the discussion in the previous subsection, features are observed in the related obvious manifestation cases of $n_f = 4$, 5 numbers of active quark flavors. The corresponding curves are presented in Fig. 2(a,b) and Fig. 3(a,b) respectively.

In general, at $n_f = 4, 5$ the character of the behavior of the Adler function and the factor $\exp(-\Delta/2)$ are similar to the one presented in Fig. 1. However, now the scale dependence of $D(Q^2)$ in the PMC/BLM approach in the N$^3$LO approximation is more noticeable than in the case of $n_f = 3$.

Why this is the case is still not clear to us. Among possible explanations are the manifestations of the effects, related to the regions where the uncertainties of the matching conditions may be detected.

### VI. CONCLUSION

In this work, we have considered the RG relation between the $e^+e^-$ annihilation Adler function $D(L, a_s)$, the photon vacuum polarization function $\Pi(L, a_s)$, and its anomalous dimension $\gamma(a_s)$ in QCD. We have provided arguments in favor of the necessity of the $\{\beta\}$ expansion of PT series for $\gamma(a_s)$ and $\Pi(L, a_s)$ to extract the scale-invariant contributions to $D(L, a_s)$ function properly and to satisfy the fundamental renormalization principles. We have demonstrated that, unless this is done, the obtained expressions for terms $\tilde{d}_k[...]$ of the $\{\beta\}$ decomposition of the Adler function will not correspond to the well-known renormalon asymptotics for its higher-order PT coefficients and the values of terms $\tilde{d}_k[0]$, defined in such way, will not be genuinely scale invariant. We emphasize that the PMC/BLM scale setting approach will be actually implemented only after the $\{\beta\}$-decomposition procedure of the PT expressions for the photon vacuum polarization function and its anomalous dimension. Thus, the photon anomalous dimension is not the conformal contribution to the Adler function. All terms of the considered $\{\beta\}$ expansion for $\gamma(a_s)$ are defined within the decomposition procedure in powers of $\beta(a_s)/a_s$.

## ACKNOWLEDGMENTS

We are grateful to G. Cvetič and S. V. Mikhailov for useful discussions. It is a pleasure to thank M. Khellat for his active participation and contribution at the initial stage of this work. A. L. K. would like to acknowledge previous round-table online discussions with S. J. Brodsky at the final stage of Bled-2022 Workshop. The work of V. S. M. was supported by the Russian Science Foundation, Agreement No. 21-71-30003 (the study of the $\{\beta\}$-expansion problem for the photon anomalous dimension and vacuum polarization operator) and by the Ministry of Education and Science of the Russian Federation as part of the program of the Moscow Center for Fundamental and Applied Mathematics, Agreement No. 075-15-2022-284 (the numerical analysis of the PMC/BLM scale setting procedure).

## APPENDIX A

The solution of the RG equation (6) for the photon vacuum polarization function can be found perturbatively, and at the $\mathcal{O}(a_s^4)$ level it has the following form for both the NS and SI contributions:

$$\Pi^{\mathrm{NS}}(L, a_s) = \Pi_0 + \gamma_0 L + (\Pi_1 + \gamma_1 L)a_s(\mu^2) + \left(\Pi_2 + (\gamma_2 + \beta_0 \Pi_1)L + \frac{1}{2}\beta_0 \gamma_1 L^2\right)a_s^2(\mu^2)$$

$$+ \left(\Pi_3 + (\gamma_3 + \beta_1 \Pi_1 + 2\beta_0 \Pi_2)L + \left(\beta_0 \gamma_2 + \frac{1}{2}\beta_1 \gamma_1 + \beta_0^2 \Pi_1\right)L^2 + \frac{1}{3}\beta_0^2 \gamma_1 L^3\right)a_s^3(\mu^2)$$

$$+ \left(\Pi_4 + (\gamma_4 + \beta_2 \Pi_1 + 2\beta_1 \Pi_2 + 3\beta_0 \Pi_3)L + \left(\beta_1 \gamma_2 + \frac{1}{2}\beta_2 \gamma_1 + \frac{3}{2}\beta_0 \gamma_3 + \frac{5}{2}\beta_0 \beta_1 \Pi_1 + 3\beta_0^2 \Pi_2\right)L^2\right.$$

$$\left. + \left(\frac{5}{6}\beta_0 \beta_1 \gamma_1 + \beta_0^2 \gamma_2 + \beta_0^3 \Pi_1\right)L^3 + \frac{1}{4}\beta_0^3 \gamma_1 L^4\right)a_s^4(\mu^2) + \ldots, \tag{A1a}$$

$$\Pi^{\mathrm{SI}}(L, a_s) = (\Pi_3^{\mathrm{SI}} + \gamma_3^{\mathrm{SI}} L)a_s^3(\mu^2) + \left(\Pi_4^{\mathrm{SI}} + (\gamma_4^{\mathrm{SI}} + 3\beta_0 \Pi_3^{\mathrm{SI}})L + \frac{3}{2}\beta_0 \gamma_3^{\mathrm{SI}} L^2\right)a_s^4(\mu^2) + \ldots. \tag{A1b}$$

The explicit solution of the RG equation (14), expressed in terms of the PT coefficients of the photon vacuum polarization function, its anomalous dimension, and the RG $\beta$ function, reads

$$D^{\mathrm{NS}}(L, a_s) = \gamma_0 + \gamma_1 a_s(\mu^2) + (\gamma_2 + \beta_0 \Pi_1 + \beta_0 \gamma_1 L)a_s^2(\mu^2)$$

$$+ (\gamma_3 + \beta_1 \Pi_1 + 2\beta_0 \Pi_2 + (\beta_1 \gamma_1 + 2\beta_0 \gamma_2 + 2\beta_0^2 \Pi_1)L + \beta_0^2 \gamma_1 L^2)a_s^3(\mu^2)$$

$$+ \left(\gamma_4 + \beta_2 \Pi_1 + 2\beta_1 \Pi_2 + 3\beta_0 \Pi_3 + (2\beta_1 \gamma_2 + \beta_2 \gamma_1 + 3\beta_0 \gamma_3 + 5\beta_0 \beta_1 \Pi_1 + 6\beta_0^2 \Pi_2)L\right.$$

$$\left. + \left(3\beta_0^2 \gamma_2 + \frac{5}{2}\beta_0 \beta_1 \gamma_1 + 3\beta_0^3 \Pi_1\right)L^2 + \beta_0^3 \gamma_1 L^3\right)a_s^4(\mu^2) + \ldots, \tag{A2a}$$

$$D^{\mathrm{SI}}(L, a_s) = \gamma_3^{\mathrm{SI}} a_s^3(\mu^2) + (\gamma_4^{\mathrm{SI}} + 3\beta_0 \Pi_3^{\mathrm{SI}} + 3\beta_0 \gamma_3^{\mathrm{SI}} L)a_s^4(\mu^2) + \ldots. \tag{A2b}$$

## APPENDIX B

### 1. Coefficients $d_k[...]$

Application of the $\{\beta\}$-decomposition procedure (19) to the PT series for the Adler function enables one to obtain the expressions for terms $d_k[...]$ and $d_k^{SI}[...]$ in relations (18a)–(18f). Within this procedure, these terms were defined previously in Refs. [28,29]. The scale-invariant contributions $d_k[0]$ and $d_k^{SI}[0]$ satisfy the relations

$$d_k[0] = \gamma_k[0], \qquad d_k^{SI}[0] = \gamma_k^{SI}[0], \tag{B1a}$$

and terms $\gamma_k[0]$, $\gamma_k^{SI}[0]$ were fixed in Eqs. (23a), (23c), and (23f)–(23h).

The remaining scale-noninvariant contributions to the $D(Q^2)$ function have the following analytical form:

$$d_2[1] = d_3[0,1] = d_4[0,0,1] = \boxed{\left(\frac{33}{8} - 3\zeta_3\right)C_F}, \tag{B1b}$$

$$d_3[1] = d_4[0,1] = \left(-\frac{111}{64} - 12\zeta_3 + 15\zeta_5\right)C_F^2 + \left(\frac{83}{32} + \frac{5}{4}\zeta_3 - \frac{5}{2}\zeta_5\right)C_F C_A, \tag{B1c}$$

$$d_3[2] = \frac{1}{2}d_4[1,1] = \boxed{\left(\frac{151}{6} - 19\zeta_3\right)C_F}, \tag{B1d}$$

$$d_4[1] = \left(-\frac{785}{128} - \frac{9}{16}\zeta_3 + \frac{165}{2}\zeta_5 - \frac{315}{4}\zeta_7\right)C_F^3 + \left(-\frac{3737}{144} + \frac{3433}{64}\zeta_3 - \frac{99}{4}\zeta_3^2 - \frac{615}{16}\zeta_5 + \frac{315}{8}\zeta_7\right)C_F^2 C_A$$
$$+ \left(-\frac{2695}{384} - \frac{1987}{64}\zeta_3 + \frac{99}{4}\zeta_3^2 + \frac{175}{32}\zeta_5 - \frac{105}{16}\zeta_7\right)C_F C_A^2, \tag{B1e}$$

$$d_4[2] = \left(-\frac{4159}{384} - \frac{2997}{16}\zeta_3 + 27\zeta_3^2 + \frac{375}{2}\zeta_5\right)C_F^2 + \left(\frac{14615}{256} + \frac{39}{16}\zeta_3 - \frac{9}{2}\zeta_3^2 - \frac{185}{4}\zeta_5\right)C_F C_A, \tag{B1f}$$

$$d_4[3] = \boxed{\left(\frac{6131}{36} - \frac{203}{2}\zeta_3 - 45\zeta_5\right)C_F}, \tag{B1g}$$

$$d_4^{SI}[1] = \left(\frac{149}{192} - \frac{39}{32}\zeta_3 + \frac{15}{16}\zeta_5 - \frac{3}{8}\zeta_3^2\right)\frac{d^{abc}d^{abc}}{d_R}. \tag{B1h}$$

The boxed analytical expressions are results for the leading renormalon-chain contributions, obtained in Ref. [62] and also presented in Ref. [7].

### 2. Coefficients $\tilde{d}_k[...]$

In the case where the photon vacuum polarization function and its anomalous dimension are not $\{\beta\}$ decomposed, the counterparts of the expressions (23a), (23c), (23f)–(23h), and (B1b)–(B1h) were obtained in Refs. [23,31,40–42] and read as follows:

$$\tilde{d}_1[0] = \gamma_1 = \frac{3}{4}C_F, \tag{B2a}$$

$$\tilde{d}_2[0] = \gamma_2 = -\frac{3}{32}C_F^2 + \frac{133}{192}C_F C_A \boxed{-\frac{11}{48}C_F T_F n_f}, \tag{B2b}$$

$$\tilde{d}_2[1] = \tilde{d}_3[0,1] = \tilde{d}_4[0,0,1] = \Pi_1 = \boxed{\left(\frac{55}{16} - 3\zeta_3\right)C_F}, \tag{B2c}$$

$$\tilde{d}_3[0] = \gamma_3 = -\frac{69}{128}C_F^3 + \left(\frac{215}{288} - \frac{11}{24}\zeta_3\right)C_F^2 C_A + \left(\frac{5815}{20736} + \frac{11}{24}\zeta_3\right)C_F C_A^2$$

$$- \left(\frac{169}{288} - \frac{11}{12}\zeta_3\right)C_F^2 T_F n_f - \left(\frac{769}{5184} + \frac{11}{12}\zeta_3\right)C_F C_A T_F n_f \boxed{-\frac{77}{1296}C_F T_F^2 n_f^2}, \quad \text{(B2d)}$$

$$\tilde{d}_3[1] = \tilde{d}_4[0,1] = \Pi_2 = \left(-\frac{143}{96} - \frac{37}{8}\zeta_3 + \frac{15}{2}\zeta_5\right)C_F^2 + \left(\frac{44215}{3456} - \frac{227}{24}\zeta_3 - \frac{5}{4}\zeta_5\right)C_F C_A$$

$$\boxed{-\left(\frac{3701}{864} - \frac{19}{6}\zeta_3\right)C_F T_F n_f}, \quad \text{(B2e)}$$

$$\tilde{d}_3[2] = \tilde{d}_4[2] = \tilde{d}_4[3] = 0, \quad \text{(B2f)}$$

$$\tilde{d}_4[0] = \gamma_4 = \left(\frac{4157}{2048} + \frac{3}{8}\zeta_3\right)C_F^4 - \left(\frac{7755}{1024} + \frac{71}{16}\zeta_3 - \frac{935}{128}\zeta_5\right)C_F^3 C_A + \left(\frac{882893}{110592} + \frac{11501}{4608}\zeta_3 + \frac{121}{256}\zeta_4 - \frac{2145}{256}\zeta_5\right)C_F^2 C_A^2$$

$$- \left(\frac{1192475}{663552} - \frac{5609}{4608}\zeta_3 + \frac{121}{256}\zeta_4 - \frac{825}{512}\zeta_5\right)C_F C_A^3 + \left(\frac{2509}{1536} + \frac{67}{32}\zeta_3 - \frac{145}{32}\zeta_5\right)C_F^3 T_F n_f$$

$$- \left(\frac{66451}{18432} - \frac{2263}{1152}\zeta_3 + \frac{143}{128}\zeta_4 - \frac{255}{64}\zeta_5\right)C_F^2 C_A T_F n_f + \left(\frac{22423}{41472} - \frac{9425}{2304}\zeta_3 + \frac{143}{128}\zeta_4 + \frac{45}{128}\zeta_5\right)C_F C_A^2 T_F n_f$$

$$+ \left(\frac{4961}{13824} - \frac{119}{144}\zeta_3 + \frac{11}{32}\zeta_4\right)C_F^2 T_F^2 n_f^2 - \left(\frac{8191}{41472} - \frac{563}{576}\zeta_3 + \frac{11}{32}\zeta_4\right)C_F C_A T_F^2 n_f^2$$

$$\boxed{+\left(\frac{107}{10368} + \frac{1}{72}\zeta_3\right)C_F T_F^3 n_f^3} + \left(\frac{3}{16} - \frac{1}{4}\zeta_3 - \frac{5}{4}\zeta_5\right)\frac{d_F^{abcd} d_A^{abcd}}{d_R} - \left(\frac{13}{16} + \zeta_3 - \frac{5}{2}\zeta_5\right)\frac{d_F^{abcd} d_F^{abcd}}{d_R} n_f, \quad \text{(B2g)}$$

$$\tilde{d}_4[1] = \Pi_3 = \left(-\frac{31}{256} + \frac{39}{32}\zeta_3 + \frac{735}{32}\zeta_5 - \frac{105}{4}\zeta_7\right)C_F^3 - \left(\frac{382033}{27648} + \frac{46219}{1152}\zeta_3 + \frac{11}{64}\zeta_4 - \frac{9305}{192}\zeta_5 - \frac{105}{8}\zeta_7\right)C_F^2 C_A$$

$$+ \left(\frac{34499767}{497664} - \frac{147473}{3456}\zeta_3 + \frac{55}{8}\zeta_3^2 + \frac{11}{64}\zeta_4 - \frac{28295}{1152}\zeta_5 - \frac{35}{16}\zeta_7\right)C_F C_A^2$$

$$- \left(\frac{7505}{13824} - \frac{1553}{72}\zeta_3 + 3\zeta_3^2 - \frac{11}{32}\zeta_4 + \frac{125}{6}\zeta_5\right)C_F^2 T_F n_f$$

$$- \left(\frac{5559937}{124416} - \frac{41575}{1728}\zeta_3 - \frac{1}{2}\zeta_3^2 + \frac{11}{32}\zeta_4 - \frac{515}{36}\zeta_5\right)C_F C_A T_F n_f$$

$$\boxed{+\left(\frac{196513}{31104} - \frac{809}{216}\zeta_3 - \frac{5}{3}\zeta_5\right)C_F T_F^2 n_f^2}, \quad \text{(B2h)}$$

$$\tilde{d}_3^{\text{SI}}[0] = \gamma_3^{\text{SI}} = \left(\frac{11}{192} - \frac{1}{8}\zeta_3\right)\frac{d^{abc} d^{abc}}{d_R}, \quad \text{(B2i)}$$

$$\tilde{d}_4^{\mathrm{SI}}[0] = \gamma_4^{\mathrm{SI}} = \left(\left(-\frac{13}{64} - \frac{1}{4}\zeta_3 + \frac{5}{8}\zeta_5\right)C_F + \left(\frac{1015}{3072} - \frac{659}{1024}\zeta_3 + \frac{33}{256}\zeta_4 + \frac{15}{256}\zeta_5\right)C_A \right. \tag{B2j}$$

$$\left. + \left(-\frac{55}{768} + \frac{41}{256}\zeta_3 - \frac{3}{64}\zeta_4 - \frac{5}{64}\zeta_5\right)T_F n_f\right)\frac{d^{abc}d^{abc}}{d_R}, \tag{B2k}$$

$$\tilde{d}_4^{\mathrm{SI}}[1] = \Pi_3^{\mathrm{SI}} = \left(\frac{431}{2304} - \frac{63}{256}\zeta_3 - \frac{1}{8}\zeta_3^2 - \frac{3}{64}\zeta_4 + \frac{15}{64}\zeta_5\right)\frac{d^{abc}d^{abc}}{d_R}, \tag{B2l}$$

where $d_3^{\mathrm{SI}} = \tilde{d}_3^{\mathrm{SI}}[0]$ and $d_4^{\mathrm{SI}} = \tilde{d}_4^{\mathrm{SI}}[0] + 3\beta_0 \tilde{d}_4^{\mathrm{SI}}[1]$. The corresponding combination of the single, double, and triple boxed analytical expressions coincide with the boxed analytical expressions of Eqs. (B1b), (B1d), and (B1g), which are related to leading renormalon-chain contributions. The second terms from these pairs of boxed terms were absorbed in unproperly defined PMC scales, while the remaining renormalon-chain effects are still contributing to the non-beta-expanded anomalous dimension.